\documentclass[structabstract]{aa}  
\usepackage{graphicx}
\usepackage{txfonts}
\usepackage{natbib}
\usepackage{longtable}
\begin{document}
   \title{Characteristics of solar-like oscillations in red giants\\ observed in the CoRoT\thanks{The CoRoT space mission which was developed and is operated by the French space agency CNES, with participation of ESA's RSSD and Science Programmes, Austria, Belgium, Brazil, Germany, and Spain. Light curves can be retrieved from the CoRoT archive: http://idoc-corot.ias.u-psud.fr.} exoplanet field}


   \author{S. Hekker\inst{1}\fnmsep\inst{2}\fnmsep\inst{3}  \and T. Kallinger\inst{4} \and F. Baudin\inst{5} \and J. De Ridder\inst{2} \and C. Barban
   \inst{6} \and F. Carrier\inst{2} \and A.P.  Hatzes\inst{7} \and W.W. Weiss\inst{4} \and A. Baglin\inst{6}}

   \institute{University of Birmingham, School of Physics and Astronomy, Edgbaston, Birmingham B15 2TT, United Kingdom\\
                 \email{saskia@bison.ph.bham.ac.uk}
           \and
                 Instituut voor Sterrenkunde, K.U. Leuven, Celestijnenlaan 200D, 3001 Leuven, Belgium
            \and
                 Royal Observatory of Belgium, Ringlaan 3, 1180 Brussels, Belgium
            \and
                 Institute for Astronomy, University of Vienna, T\"urkenschanzstrasse 17, A-1180 Vienna
            \and
                 Institute d'Astrophysique Spatiale, UMR 8617, Universit\'e Paris XI, B\^atiment 121, 91405 Orsay Cedex, France
            \and
                 LESIA, UMR8109, Universit\'e Pierre et Marie Curie, Universit\'e Denis Diderot, Observatoire de Paris, 92195 Meudon Cedex, France
            \and
                 Th\"uringer Landessternwarte, D-07778 Tautenburg, Germany
             }

   \date{Received ; accepted }

 
  \abstract
   {Observations during the first long run ($\sim$150 days) in the exo-planet field of CoRoT increase the number of G-K giant stars for which solar-like oscillations are observed by a factor of 100. This opens the possibility to study the characteristics of their oscillations in a statistical sense.}
   {We aim to understand the statistical distribution of the frequencies of maximum oscillation power ($\nu_{max}$) in red giants and to search for a possible correlation between $\nu_{max}$ and the large separation ($\Delta \nu$).}
   {Red giants with detectable solar-like oscillations are identified using both semi-automatic and manual procedures. For these stars, we determine $\nu_{max}$ as the centre of a Gaussian fit to the oscillation power excess. For the determination of $\Delta \nu$, we use the autocorrelation of the Fourier spectra, the comb response function and the power spectrum of the power spectrum.}
   {The resulting $\nu_{max}$ distribution shows a pronounced peak between 20 - 40 $\mu$Hz.  For about half of the stars we obtain $\Delta \nu$ with at least two methods. The correlation between $\nu_{max}$ and $\Delta \nu$ follows the same scaling relation as inferred for solar-like stars.}
   {The shape of the $\nu_{max}$ distribution can partly be explained by granulation at low frequencies and by white noise at high frequencies, but the population density of the observed stars turns out to be also an important factor. From the fact that the correlation between $\Delta \nu$ and $\nu_{max}$ for red giants follows the same scaling relation as obtained for sun-like stars, we conclude that the sound travel time over the pressure scale height of the atmosphere scales with the sound travel time through the whole star irrespective of evolution. The fraction of stars for which we determine $\Delta \nu$ does not correlate with $\nu_{max}$ in the investigated frequency range, which confirms theoretical predictions. }

   \keywords{stars: red giants -- stars: oscillations -- methods: observational -- techniques: photometric }

   \maketitle
%

\section{Introduction}
Before the CoRoT era, the presence of solar-like oscillations was firmly established for a few red (G-K) giant stars only. We refer to \citet{deridder2009} for an overview of these results. With this low number of firm detections, different authors reached different conclusions in terms of the presence of only radial modes with short lifetimes or radial and non-radial oscillation modes with much longer lifetimes.

Observations with CoRoT increased the number of giants with a clear detection of solar-like oscillations by almost a factor of 100. Using these observations, \citet{deridder2009} recently presented the discovery of non-radial oscillations modes with long lifetimes ($\geq$ 50 days). 
This result is important for red giant seismology, not only because the frequencies of modes with long lifetimes can be more precisely determined, but also because non-radial modes contain more information about the internal structure of giants than radial modes only.

Subsequently, \citet{kallinger2009} derived for 31 giants in the sample of \citet{deridder2009}, the frequency of maximum power and the large separation, and used these values to estimate the mass and  
the radius.

In the present paper, we exploit the fact that we now have, for the first time, a large sample of red giants with established solar-like oscillations, which opens up the possibility to analyse the characteristics of the power spectrum in a statistical way. In particular, we investigate the distribution of the frequency of maximum oscillation power $\nu_{max}$, and the large frequency separations $\Delta\nu$, i.e. the frequency difference between modes with the same degree and consecutive orders. These parameters change with stellar age, and their histogram can therefore provide insight into the population of observed giants. 





\begin{figure}
\begin{minipage}{\linewidth}
\centering
\includegraphics[width=\linewidth]{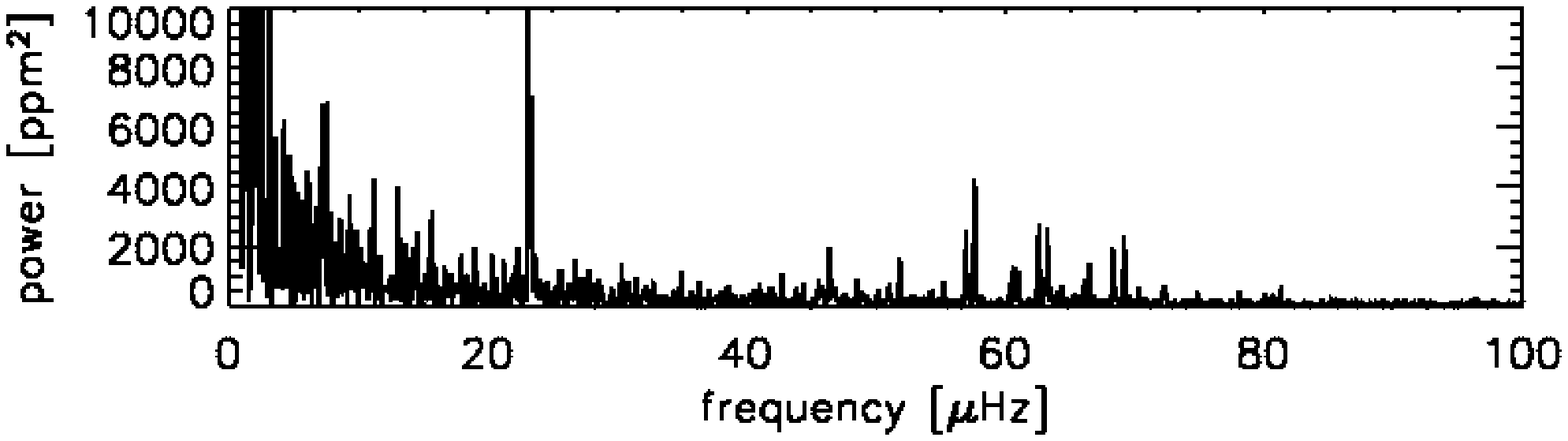}
\end{minipage}
\begin{minipage}{\linewidth}
\centering
\includegraphics[width=\linewidth]{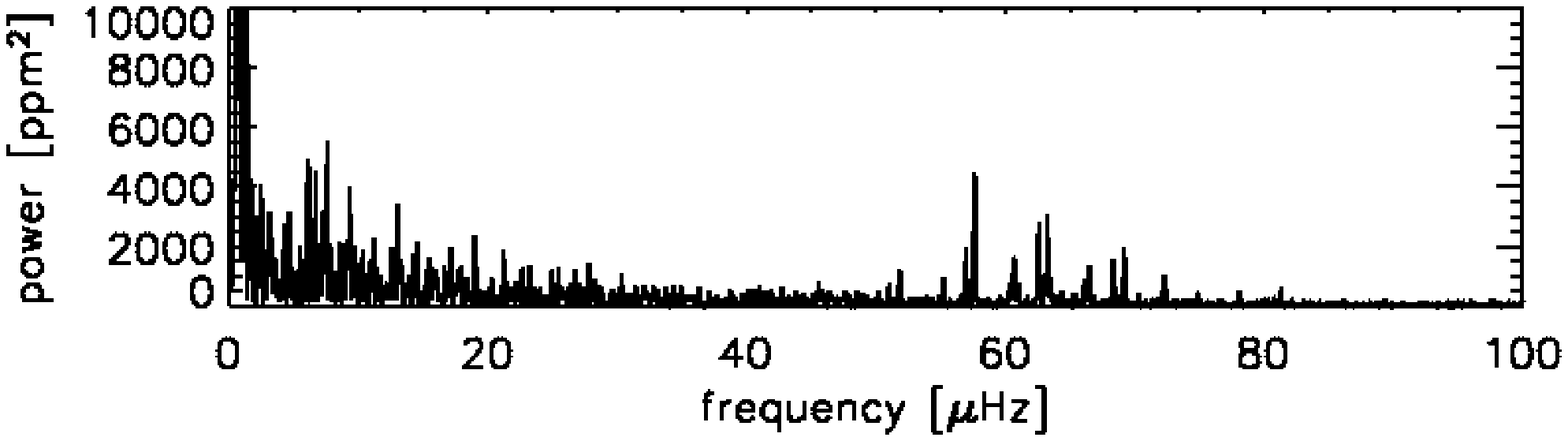}
\end{minipage}
\caption{Uncorrected (top) and corrected (bottom) power spectrum of the star with CoRoT-ID 100697490. The peak at about 23 $\mu$Hz is twice the daily frequency.}
\label{coruncor}
\end{figure}

\begin{figure}
\centering
\includegraphics[width=\linewidth]{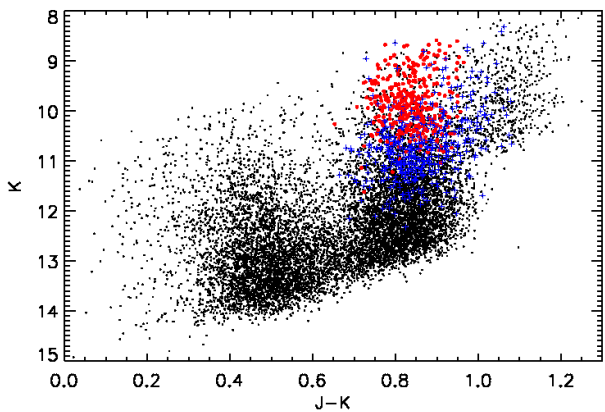}
\caption{ Colour-magnitude diagram of the observed field stars using the J and K photometric passbands \citep{deleuil2006}. The red dots represent the semi-automatic selected red giants and the blue crosses the ones selected by eye only. The black dots represent all other targets in the field. Note that these photometric data have not been dereddened. }
\label{colour}
\end{figure}

\section{Data}
The CoRoT data used in this work are the reduced (N2) monochromatic light curves \citep{auvergne2009} provided by the CoRoT data centre of the first long run (LRc01) of about 150 days from May to October 2007, when the satellite was pointed towards the galactic centre (($\alpha$, $\delta$)=(290.89$^{\circ}$,0.46$^{\circ}$)). For information on the CoRoT data reduction, we refer to \citet{baglin2006cdc}. The data used for the present investigation are obtained in the so-called exofield. The light curves consist of approximately 330\,000 points with a typical time step of 32 s. Although the light curves of some of the targets of interest contain fewer data points, with a cadence of 512 s. In all data sets, we removed all points that were flagged as unreliable. 

Many of the light curves show signs of instrumental effects. A proper treatment of these effects would require an in-depth knowledge of the instrument and is currently under investigation by the CoRoT data  
centre. For the purpose of this paper it suffices to mitigate the instrumental effects as follows. First, we eliminated possible trends by fitting and subtracting a second-order polynomial. Secondly, the  
occasional jumps due to high-energy particles were first detected by comparing flux-levels in consecutive time intervals at least 10 times larger than the expected oscillation periods, and were then removed by fitting and subtracting a polynomial background on each side of the jump. Finally, outliers were removed using a 4-sigma clipping around the mean flux value. Comparing the power spectra of the corrected time series with power spectra of the uncorrected time series, revealed that the instrumental effects are not dominant, see Fig.~\ref{coruncor} for an example. To see the effect at low frequencies, we looked at the power spectra of B stars, for which the noise at low frequencies should be dominated by instrumental noise, as there is no surface granulation for these stars. Assuming that the instrumental noise for B stars is roughly the same as for giants,  we could conclude that the low-frequency noise in the red giant power spectra is not dominated by instrumental noise, with only a few exceptions in cases with a large number of jumps.

\begin{figure}
\centering
\includegraphics[width=\linewidth]{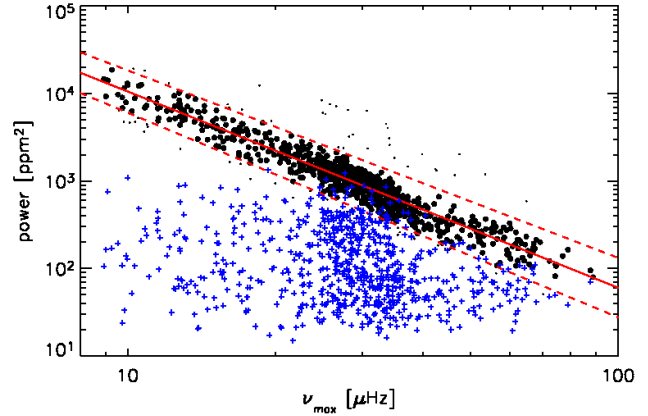}
\caption{The height of the oscillation power excess ($D$ in Eq.~\ref{powerlaw}) is shown as a function of $\nu_{max}$ with black dots (outliers are indicated with smaller dots). The solid red line represents the linear fit in log space $D$ = 6.3$-$2.2$\nu_{max}$, with the 3$\sigma$ interval indicated with red dashed lines. The blue crosses represent white noise ($E$ in Eq.~\ref{powerlaw}, further discussed in Section 4.1).}
\label{gaussnumax}
\end{figure}

\section{Identification of oscillations in red giants}
Visual and near-IR photometry are available for all stars observed in the exo-field during run LRc01 \citep{deleuil2006}. These colours are affected by reddening, but as shown by \citet{bessel1988} near-IR colours are least affected, and provide a first estimate of the spectral type.  A J-K versus K colour-magnitude diagram is shown in Fig.~\ref{colour}. 

Red giants with solar-like oscillations cannot (yet) be classified with the automated supervised classification algorithm developed for CoRoT \citep{debosscher2007}. This is mainly due to the low amplitudes of these oscillations and the low number of detections prior to CoRoT observations. These low numbers hamper a reliable class definition, which is needed by the classification algorithm used by \citet{debosscher2007}.

We used the same semi-automatic procedure as described by \citet{deridder2009} to select red-giant stars for which we can detect solar-like oscillations. In addition, for reasons explained below, we also inspected the Fourier spectra by eye to check for oscillation features in stars not selected with the semi-automatic procedures.  In the selection by eye we inspected the power spectrum between 0 and 120 $\mu$Hz for broad power excess (in case of modes with short lifetimes) or a cluster of several individual frequency peaks at intervals of a few $\mu$Hz (modes with long lifetimes). At $\sim$163 $\mu$Hz a strong frequency peak due to the orbital period of CoRoT is present which has sidelobes at intervals of 11.57 $\mu$Hz down to about 120 $\mu$Hz. These features in the power spectrum limit the frequency range for which we search for oscillation signatures.

To validate our selection we fitted a model to the smoothed power spectrum consisting of a power law and a Gaussian representing the oscillation power excess respectively, see Eq.~\ref{powerlaw}, with $\nu$ the frequency, $A$ the amplitude of the background, $B$ the characteristic timescale, $C$ the slope of the power law, $D$ the height of the oscillation power excess, $\nu_{max}$ the frequency of maximum oscillation power, $\sigma$ the width of the oscillation excess and $E$ the white noise. The smoothing is performed using a moving average with a varying width of 4 times the expected large separation at each frequency \citep{kjeldsen2008}. See Section 4 for the correlation between $\Delta \nu$ and $\nu_{max}$. We take frequency changes in the oscillation spectrum into account in the smoothing to pursue a homogeneous analyses for all stars in the sample.
\begin{equation}
P(\nu)=\frac{A}{(1+(B \cdot \nu)^C)} +D \cdot e^{\frac{(\nu_{max} - \nu)^2}{2 \sigma^2}} + E,
\label{powerlaw}
\end{equation}
This fitting procedure is a simplification of the background fitting used by \citet{aigrain2004}, which is sufficient as we only use it to validate the presence of the oscillations and pinpoint their frequency of maximum oscillation power (Section 4.1). 

For the validation of our candidates, we first removed stars with non-converging or spurious fits from the sample. Then we looked into the fitted height of the power excess ($D$ in Eq.\ref{powerlaw}).
\citet{chaplin2009} recently presented a scaling relation for mode lifetimes and they find that the maximum mode height of the oscillations ($H$)  depends predominantly on the surface gravity of stars ($g$), i.e., $H \sim g^{-2}$. Furthermore, it is known that $\nu_{max} \sim \nu_{ac} \sim gT^{-1/2}_{\rm eff}$, with $\nu_{ac}$ the acoustic cut off frequency \citep{brown1991,kjeldsen1995}. Combining these 2 relations, in which we approximate the effective temperature ($T_{\rm eff}$) to be constant, we expect that $H \sim \nu_{max}^{-2}$ and thus also that the height of the Gaussian fit to the oscillations power excess decreases with increasing $\nu_{max}$.
In Fig.~\ref{gaussnumax}, we indeed see the expected trend and the best fit provides us with an exponent of -2.2 instead of -2, which is predicted by the scaling relation. This may be due to the fact that the scaling relation for $H$ is based on narrow band photometry, while we use the height of the power excess in broadband photometry. 

As we expect that the scaling relations should be valid for all stars at hand, we exclude stars with fit parameters which fall outside the 3$\sigma$ interval around the correlation between the fitted height of the oscillation excess and the frequency of maximum oscillation power. This left us with 778 oscillating red giant candidates, which are indicated in Fig.~\ref{colour} and in the online Table~1. 

\section{Characteristics of solar-like oscillation in red giants}

\begin{figure}
\centering
\includegraphics[width=\linewidth]{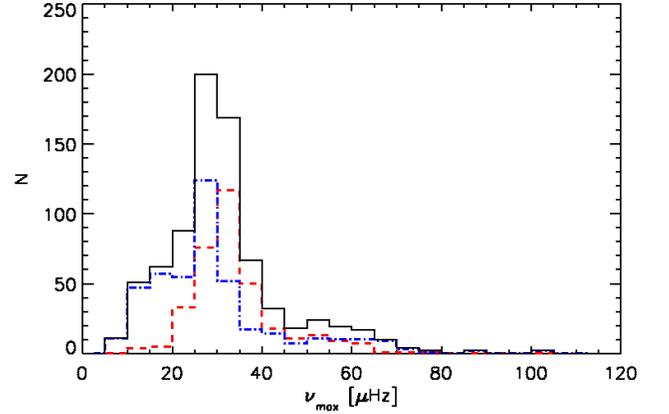}
\caption{The histogram of the frequencies at maximum oscillation power for all oscillating red-giant candidates is shown in black. The red dashed and blue dash-dot histograms show the $\nu_{max} $ distribution for oscillating red giants selected with the semi-automatic procedure and the ones selected manually, respectively.} 
\label{numaxhisto}
\end{figure}

\subsection{Frequency of maximum oscillation power}
The frequency of maximum oscillation power, defined as the centre of the Gaussian fitted to the oscillation power excess, is the first parameter of interest we investigate here. The $\nu_{max}$ distribution is plotted in Fig.~\ref{numaxhisto}. This distribution shows a clear maximum between 20-40 $\mu$Hz. Before interpreting this distribution, we first investigate possible selection / observational biases that might influence it.

We first investigate possible selection effects imposed by the semi-automatic procedure. Therefore we inspected all Fourier spectra by eye. We checked the Fourier spectra for excess power, as explained above, and were able to identify additional stars with power excess (blue symbols in Fig.~\ref{colour}). We identified more stars with oscillations at lower frequencies, where our semi-automatic procedure was truncated because of possible contamination with power excess due to granulation. In general these additional stars have a similar distribution of $\nu_{max}$ as the ones selected with the semi-automatic procedure, see Fig.~\ref{numaxhisto}. Based on this, we discard the possibility that the peaked distribution is due to the identification method we used to select red giants with power excess due to solar-like oscillations.


In terms of observational biases, it is known that granulation is present in red-giant stars with power at low frequencies in the Fourier spectrum. In addition to the increase of granulation power, the width $w$ of the oscillation power excess decreases with decreasing $\nu_{max}$, as this scales as $ w \sim |\nu_{max}-\nu_{ac}|$ \citep{kjeldsen1995}. Both the increase of granulation and the smaller width of the power excess make it increasingly difficult to detect oscillations at low frequencies. Therefore the number of stars at low $\nu_{max}$ is most likely underestimated.  

\begin{figure}
\centering
\includegraphics[width=\linewidth]{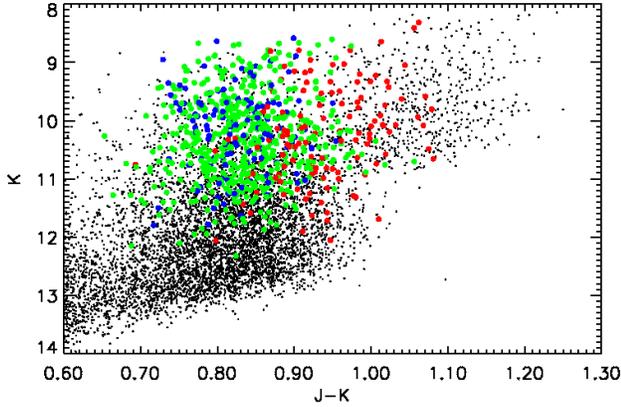}
\caption{The red giant branch of the colour-magnitude diagram as shown in Fig.~\ref{colour} The black dots represent all targets in the field. Stars with $\nu_{max} < 20 \mu$Hz, 20 $\mu$Hz $\leq \nu_{max} \leq 40 \mu$Hz, $\nu_{max} > 40 \mu$Hz are indicated with red, green and blue dots respectively.}
\label{colournumax}
\end{figure}

To further investigate the low number of stars with $\nu_{max} <$ 20 $\mu$Hz, we checked whether stars with $\nu_{max} <$ 20 $\mu$Hz lay in a specific part of the colour-magnitude diagram. In Fig.~\ref{colournumax}, the red-giant branch is shown with the stars for which we obtained $\nu_{max}$ indicated in red, green and blue for stars with $\nu_{max} < 20$ $\mu$Hz, stars with 20 $\mu$Hz $\leq \nu_{max} \leq$ 40 $\mu$Hz and stars with $\nu_{max} >$ 40 $\mu$Hz, respectively. Clearly, the stars with lowest $\nu_{max}$ appear in the reddest and brightest part of the colour-magnitude diagram. This part of the colour-magnitude diagram is also less well populated, which decreases the probability of observing these stars. The low number of stars with $\nu_{max}$ $<$ 20 $\mu$Hz can therefore be explained by detection difficulties due to both granulation and decreasing width of the power excess, and low population density of stars with oscillations in this frequency range.

The decrease in the number of stars with $\nu_{max}$ at frequencies $>$ 40 $\mu$Hz is partly caused by the fact that at higher frequencies the height of the power excess decreases, as can be seen in Fig.~\ref{gaussnumax}, and as is predicted by the scaling relations of \citet{kjeldsen1995}. When we look at the white noise ($E$ in Eq.~\ref{powerlaw}), whose values are shown as blue crosses in Fig.~\ref{gaussnumax}, we indeed see that at $\nu_{max} \sim$ 40 $\mu$Hz the height of the oscillation power excess starts to decrease below the noise level of some stars with oscillations at lower frequencies. So for stars with $\nu_{max}$ $>$ 40 $\mu$Hz we can only detect solar-like oscillations for stars with low noise levels. 

Naively, one would expect that the number of stars with a noise level below 100~ppm$^2$ would not depend on the frequency of maximum oscillation power. This would imply that we would be able to detect oscillations in a similar number of stars with these low noise levels irrespective of $\nu_{max}$.Interestingly, the density of stars with a white noise level $\leq$ 100~ppm$^2$ is much larger in the range 20 $\mu$Hz $\leq \nu_{max} \leq$ 40 $\mu$Hz than at higher $\nu_{max}$ values, see the blue crosses in Fig.~\ref{gaussnumax}. This might again indicate that the population of observed stars with $\nu_{max} >$ 40 $\mu$Hz is smaller than for stars with 20 $\mu$Hz $\leq \nu_{max} \leq$ 40 $\mu$Hz. 

Population synthesis simulations to study the probability of observing a star at  a certain position in the HR diagram are performed by \citet{miglio2009}. In their study, \citet{miglio2009} simulate the composite stellar population of the observed field, using a population synthesis code that takes into account the morphology of the galaxy and star formation history. For further details we refer to \citet{miglio2009} and references therein. The results for $\nu_{max}$ of these population synthesis simulations are in agreement with the observations presented here.

\begin{figure}
\centering
\includegraphics[width=\linewidth]{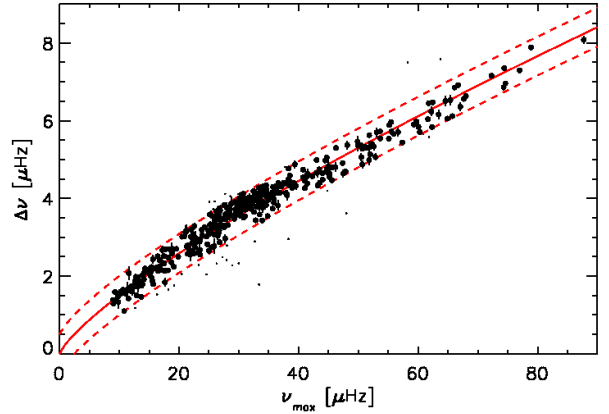}
\caption{$\Delta \nu$ as a function of $\nu_{max}$, the small dots indicate $\Delta \nu$ values with a standard deviation $\leq 0.1 \mu$Hz, but outside the 0.5 $\mu$Hz interval (red dashed lines) around the correlation $\Delta \nu = \Delta \nu_{\sun} \cdot (\nu_{max}/\nu_{max \sun})^{0.784\pm0.003}$ (red solid line). The large dots are the 367 stars with $\Delta \nu$ values along the correlation and with standard deviation between the different measures $\leq$ 0.2 $\mu$Hz }
\label{deltanumax}
\end{figure}

\subsection{Large separation}

A second interesting parameter is the large separation ($\Delta \nu$) between frequencies of modes with the same degree and consecutive overtones of the radial order.
A theoretical investigation by \citet{dupret2009} shows that the power spectra of the red-giant stars are different for different evolutionary phases. For more evolved stars only oscillations trapped in the outer cavity (p modes) can reach observable amplitudes at the surface of the star and for high-order low-degree modes these can show asymptotic, i.e., regular behaviour \citep{tassoul1980}. Other (less evolved) stars show a more dense and / or an irregular frequency pattern, which can be explained by the fact that the observed oscillations are influenced by their behaviour in both the p-mode and g-mode cavity.

Due to the orbital frequencies of CoRoT at $\sim$163 $\mu$Hz with several side lobes at 11.57 $\mu$Hz intervals we were not able to investigate the less evolved stars at frequencies 150-200 $\mu$Hz \citep[model A,][]{dupret2009}. Therefore, we expect to observe mainly more evolved stars for which theory predicts regular frequency patterns.

To study $\Delta \nu$, we first selected the frequency range in which we will compute $\Delta \nu$. This range is scaled from the Sun and defined as $\nu_{max} \pm 0.5 \cdot (\nu_{max}/\nu_{max \sun}) \cdot w_{\sun}$, with $w_{\sun}$ the width of the oscillation excess in the Sun. In this range, we compute the autocorrelation of the Fourier spectrum, the comb response function \citep{kjeldsenea1995} and the autocorrelation of the time series (which is equivalent to the power spectrum of the power spectrum) \citep{roxburgh2006}. For each of these methods we identified the highest peak as the possible $\Delta \nu$ (for the power spectrum of the power spectrum we assume $\Delta \nu$/2, see \citet{roxburgh2006}). Then we computed the average $\Delta \nu$ and its standard deviation from the three values we obtained, taking the possibility that we identified $\Delta \nu$/2 or 2$\Delta \nu$ into account. If one of the values is off by more than the standard deviation then it is excluded. 

Resulting $\Delta \nu$ values with a standard deviation $\leq$ 0.1 $\mu$Hz are considered to be more reliable and show a clear correlation with $\nu_{max}$, see Fig~\ref{deltanumax}. This correlation follows the power law $\Delta \nu = \Delta \nu_{\sun} \cdot (\nu_{max}/\nu_{max \sun})^{0.784 \pm 0.003}$, which agrees well with the predictions and results of \citet{stello2009} that have been made for solar-like main-sequence and sub-giant stars. All stars with a $\Delta \nu$ consistent with the described power law and additionally a standard deviation between the different measures $\leq$ 0.2 $\mu$Hz are considered to have oscillation frequencies that occur at regular intervals. 

\begin{figure}
\centering
\includegraphics[width=\linewidth]{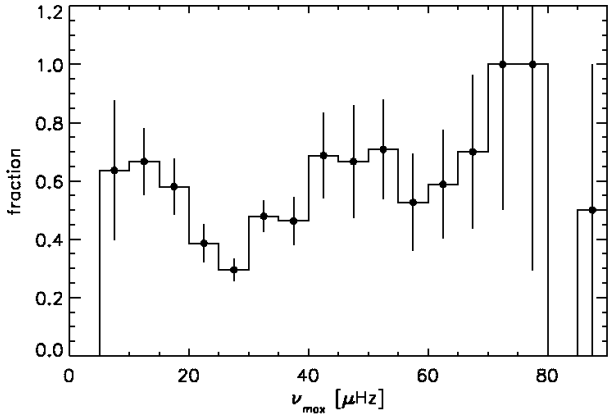}
\caption{Fraction of stars in each $\nu_{max}$ interval for which we could determine $\Delta \nu$. The error bars are calculated assuming Poisson statistics (i.e., the fraction of stars for which we obtain $\Delta \nu$ divided by the square root of the number of stars for which we obtain $\Delta \nu$). Note that the dip in this distribution between 20 and 40 $\mu$Hz coincides with most stars for which we obtained $\Delta \nu$ with at least two methods with a standard deviation $\leq$ 0.1 $\mu$Hz, but which did not follow the power law, see the small dots in Fig.~\ref{deltanumax}.}
\label{fracnumaxhisto}
\end{figure}

Fig.~\ref{fracnumaxhisto} shows the fraction of stars in each $\nu_{max}$ bin for which we could determine $\Delta \nu$ as described above.  With a two-sided Kolmogorov-Smirnov test \citep{press2002}, we find a 3\% probability that the distribution is flat, i.e., that the fraction of stars for which we compute a reliable $\Delta \nu$ is the same for all $\nu_{max}$ bins. Nevertheless, there is no correlation present between the fraction of stars for which we obtain $\Delta \nu$ and $\nu_{max}$. This is consistent with the theoretical models B, C, D and E of \citet{dupret2009}, for which regular frequency patterns are expected. The reason why we only detect $\Delta \nu$ for 367 stars might be due to noise / granulation peaks which could hamper the automatic determination. It might also be due to less well trapped $\ell$ = 1 modes as predicted by \citet{dupret2009} for stars intermediate in the red giant branch.
Furthermore, in the scaling relation used here, no dependence on stellar parameters is included.

We note here that for an additional 303 stars a value for $\Delta \nu$ within the 0.5 $\mu$Hz interval around the correlation could be obtained with only one of the methods. Including these less reliable values in the fraction of stars in each $\nu_{max}$ interval sustains the conclusion that no correlation is present between the fraction of stars for which we determine $\Delta \nu$ and $\nu_{max}$ in the frequency range considered here.

\section{Summary and Conclusions}
We were able to detect power excess resembling solar-like oscillations in 778 stars observed in the CoRoT exofield LRc01 at frequencies typical for solar-like oscillations in red (G-K) giants. These detections were made with a semi-automatic procedure with which we were able to detect oscillation power excess for the brightest stars at frequencies $>$ 20 $\mu$Hz. For fainter stars, with a less pronounced power excess, and more luminous stars, with an excess at lower frequencies, the semi-automatic procedure was less successful and an inspection by eye is performed to investigate the presence of red-giant stars with solar-like oscillations in these frequency regimes. 

This large number of detections increases the number of red-giant stars with observed solar-like oscillations by a factor of 100 and allows for a statistical investigation into the general properties of these stars. 

From the distribution of $\nu_{max}$, it becomes clear that it is most likely to observe red giants with oscillation power between 20 and 40 $\mu$Hz. Apart from the fact that we suffer from biases due to granulation / decreasing width of the oscillation excess at low frequencies and white noise at high frequencies, the increase / decrease in the number of stars with $\nu_{max}$ is also influenced by a varying population density. Evidence for this conclusion is also demonstrated by population synthesis studies, for which we refer to \citet{miglio2009}.

For about half of the stars (367) we could obtain consistent values for $\Delta \nu$ with at least two methods. These values follow the power law $\Delta \nu \sim (\nu_{max})^{0.784 \pm 0.003}$. A similar result has already been found for solar-like main-sequence and sub-giant stars \citep{stello2009}. This correlation is equivalent with $\Delta \nu \sim c_s/R \sim (c_s/H_p)^{0.784\pm0.003}$, where $c_s$ is the sound speed, $R$ the radius and $H_p$ the pressure scale height of the atmosphere \citep{kjeldsen1995}. From the fact that this relation holds for both solar-like stars and red giants, we conclude that the sound travel time over the pressure scale height of the atmosphere scales with the sound travel time through the whole star in the same way in both evolutionary states.

Furthermore, we find that the fraction of stars for which we obtain a regular spectrum varies in a complicated way with $\nu_{max}$ in the frequency range up to 100 $\mu$Hz. This is in agreement with the predictions by \citet{dupret2009} as we observe mainly stars intermediate or high in the red giant branch. For these stars modes trapped in the outer cavity can reach observational amplitudes and for high-order low-degree modes the frequencies will follow the asymptotic relation \citep{tassoul1980}.
\newline
\newline
The results obtained for red giants with CoRoT are important for improving our understanding of solar-like oscillations in red-giant stars. These observations allow us to study, for instance, the excitation and damping of these oscillations and the time scales at which these processes occur. Studying these parameters as a function of evolution will also be possible with the large number of red giants showing oscillations. Furthermore, non-radial modes with long lifetimes \citep{deridder2009} open a way to study the internal structure of individual giants in detail.  These studies are currently underway.



\begin{acknowledgements}
SH wants to thank W.J. Chaplin for useful discussions and A.-M. Broomhall for carefully reading the manuscript. SH  acknowledges financial support from the Belgian Federal Science Policy (ref: MO/33/018). TK and WWW acknowledges support by the Austrian Research Promotion Agency (FFG), and the Austria Science Fund (FWF P17580). FC is a postdoctoral fellows of the Fund for Scientific Research Flanders. APH acknowledges the support of grant 50OW0204 from the Deutsches Zentrum f\"ur Luft- und Raumfahrt e. V. (DLR). We would like to thank our referee T.R. Bedding for valuable comments, which helped to improved the manuscript considerably.
\end{acknowledgements}

\bibliographystyle{aa}
\bibliography{11858bib}

\longtab{1}{
\begin{longtable}{ccccrcc}
\caption{\label{autoselect} Red giant candidates selected in the CoRoT exofield during run LRc01: CoRoT identification number (CoRoT-ID), apparent magnitude (m$_{v}$), right ascension (ra) in degrees, declination (dec) in degrees, frequency of maximum oscillation power ($\nu_{max}$)  and large separation ($\Delta \nu$) in $\mu$Hz when obtained, see text. The $\Delta \nu$ values between () are values which follow the $\Delta \nu \sim \nu_{max}^{0.784 \pm 0.003}$ relation but are only obtained with one method. 'sap' means that a stars is selected with a semi-automatic procedure and 'man' means that the star is selected by eye only.}\\
\hline\hline
CoRoT-ID & m$_{v}$ & ra (degrees) & dec (degrees) & $\nu_{max}$ ($\mu$Hz) & $\Delta \nu$ ($\mu$Hz) & classification\\
\hline
\endfirsthead
\caption{continued.}\\
\hline\hline
CoRoT-ID & m$_{v}$ & ra (degrees) & dec (degrees) & $\nu_{max}$ ($\mu$Hz) & $\Delta \nu$ ($\mu$Hz) & selection\\
\hline
\endhead
\hline
\endfoot
100411979 & 14.875 & 290.58484 & 1.69915 & 29.7 & (3.6) & man\\
100440565 & 13.149 & 290.62716 & 1.64784 & 29.1 & 3.6 & sap\\
100475529 & 13.574 & 290.67846 & 1.78820 & 32.3 & 3.9 & sap\\
100482282 & 14.125 & 290.68833 & 1.72073 & 13.1 & 1.7 & man\\
100483847 & 12.700 & 290.69070 & 1.52023 & 39.1 & 4.4 & sap\\
100486326 & 13.465 & 290.69441 & 1.40077 & 10.0 & (1.7) & man\\
100497523 & 12.684 & 290.71066 & 1.40175 & 34.2 &  & sap\\
100500736 & 14.041 & 290.71553 & 1.37232 & 30.7 & 4.1 & sap\\
100502521 & 15.330 & 290.71812 & 1.47512 & 28.6 & (3.6) & man\\
100503016 & 14.278 & 290.71883 & 1.33593 & 10.9 & (1.5) & man\\
100503737 & 14.414 & 290.71987 & 1.42503 & 25.7 & (2.9) & man\\
100511611 & 14.340 & 290.73162 & 1.64333 & 11.5 & 1.6 & man\\
100515307 & 14.795 & 290.73734 & 1.72422 & 26.7 & 3.7 & man\\
100516923 & 14.975 & 290.73967 & 1.81010 & 45.9 & (4.7) & man\\
100516924 & 14.221 & 290.73967 & 1.34772 & 38.6 & 4.4 & man\\
100518590 & 14.590 & 290.74216 & 1.45303 & 34.0 &  & man\\
100520462 & 14.368 & 290.74493 & 1.33203 & 14.1 & 1.9 & man\\
100523200 & 13.359 & 290.74899 & 1.48020 & 15.7 & 1.8 & man\\
100525640 & 14.941 & 290.75262 & 1.59460 & 31.2 & (3.9) & sap\\
100528464 & 13.405 & 290.75684 & 1.63503 & 14.3 & 2.2 & man\\
100530452 & 13.649 & 290.75968 & 1.28913 & 35.8 & 4.0 & sap\\
100545483 & 12.724 & 290.78246 & 1.49500 & 61.6 & 6.0 & sap\\
100546715 & 15.318 & 290.78436 & 1.59082 & 28.1 & (3.3) & man\\
100552455 & 14.274 & 290.79288 & 1.51853 & 63.0 & (6.1) & man\\
100555173 & 12.394 & 290.79705 & 1.31657 & 31.4 & (3.7) & sap\\
100556001 & 12.540 & 290.79836 & 1.59940 & 45.2 & 4.6 & sap\\
100556055 & 13.199 & 290.79845 & 1.84249 & 44.8 & (4.8) & sap\\
100556225 & 14.388 & 290.79873 & 1.55524 & 27.3 & (3.6) & man\\
100562083 & 13.969 & 290.80760 & 1.82471 & 41.5 & 4.6 & man\\
100564275 & 14.494 & 290.81101 & 1.75957 & 34.4 & (4.0) & sap\\
100573220 & 13.259 & 290.82440 & 1.83611 & 33.3 & 4.3 & sap\\
100586764 & 14.155 & 290.84463 & 1.47716 & 25.2 & (3.3) & sap\\
100591536 & 14.604 & 290.85186 & 1.06328 & 35.5 & (4.4) & man\\
100591900 & 14.030 & 290.85235 & 1.63926 & 37.8 &  & man\\
100592544 & 13.690 & 290.85334 & 1.69705 & 44.0 & 4.5 & sap\\
100593268 & 13.339 & 290.85442 & 1.26642 & 29.6 & (3.3) & sap\\
100593396 & 13.554 & 290.85460 & 1.47886 & 45.6 & 4.5 & sap\\
100596299 & 14.920 & 290.85899 & 1.75909 & 40.7 & 4.5 & man\\
100597075 & 13.749 & 290.86020 & 1.29505 & 12.7 & 1.8 & man\\
100597609 & 13.024 & 290.86096 & 1.45091 & 36.6 & 3.8 & sap\\
100598266 & 14.609 & 290.86194 & 1.77280 & 9.3 & 1.6 & man\\
100602088 & 14.373 & 290.86763 & 1.02124 & 30.2 & 3.8 & sap\\
100605342 & 14.040 & 290.87237 & 1.30235 & 27.3 &  & man\\
100606331 & 13.733 & 290.87381 & 1.31963 & 23.4 & 2.9 & man\\
100614937 & 14.668 & 290.88591 & 1.47120 & 27.5 & (2.8) & man\\
100615971 & 13.160 & 290.88742 & 1.87220 & 32.5 & 3.6 & sap\\
100630893 & 12.990 & 290.90775 & 1.76520 & 27.8 & (3.4) & sap\\
100637413 & 13.150 & 290.91673 & 1.87779 & 28.3 & (3.1) & sap\\
100644163 & 13.375 & 290.92629 & 1.46415 & 38.2 & (4.6) & sap\\
100652335 & 14.547 & 290.93765 & 1.70149 & 23.4 & 2.9 & man\\
100652396 & 16.049 & 290.93775 & 1.46657 & 19.3 & (2.7) & man\\
100654821 & 12.632 & 290.94131 & 1.89830 & 31.7 & (3.8) & sap\\
100657953 & 14.573 & 290.94618 & 1.55220 & 15.2 & 2.1 & man\\
100662481 & 12.898 & 290.95388 & 1.38109 & 50.7 & 4.9 & sap\\
100667041 & 15.216 & 290.96172 & 1.19518 & 19.1 & 2.2 & man\\
100667742 & 15.506 & 290.96290 & 1.16010 & 32.7 & (4.1) & man\\
100678257 & 14.681 & 290.98050 & 1.26305 & 25.6 &  & man\\
100678505 & 12.198 & 290.98093 & 1.67693 & 64.3 & (6.4) & man\\
100679411 & 14.681 & 290.98248 & 1.29397 & 25.5 & 3.6 & man\\
100682488 & 13.691 & 290.98765 & 1.17854 & 50.5 & 5.1 & sap\\
100688417 & 14.113 & 290.99775 & 1.61537 & 19.1 & (2.5) & man\\
100688591 & 13.708 & 290.99806 & 1.54563 & 35.8 &  & sap\\
100697490 & 12.541 & 291.01319 & 1.84065 & 60.9 & (5.7) & sap\\
100698309 & 13.648 & 291.01460 & 1.54950 & 16.4 &  & man\\
100705263 & 14.241 & 291.02636 & 0.96980 & 35.6 & 4.1 & sap\\
100707942 & 14.521 & 291.03102 & 1.27638 & 62.8 & (6.3) & man\\
100714007 & 14.032 & 291.04135 & 1.83044 & 27.4 & (3.4) & sap\\
100714474 & 13.815 & 291.04208 & 1.52403 & 36.2 & 4.1 & sap\\
100716817 & 13.858 & 291.04621 & 1.63556 & 35.0 & 4.0 & sap\\
100722680 & 14.991 & 291.05657 & 1.05631 & 12.8 & 2.0 & man\\
100724563 & 13.601 & 291.06068 & 1.68904 & 10.3 & 1.5 & man\\
100725658 & 14.635 & 291.06352 & 1.77049 & 22.0 & 3.0 & man\\
100729865 & 13.806 & 291.07170 & 1.13569 & 26.1 & 3.3 & man\\
100733133 & 13.624 & 291.07624 & 0.58899 & 34.7 & (3.8) & sap\\
100733870 & 14.204 & 291.07717 & 0.88983 & 30.9 & 3.9 & sap\\
100736020 & 12.623 & 291.08004 & 1.64963 & 34.8 & 3.5 & sap\\
100736789 & 13.046 & 291.08107 & 1.19551 & 9.7 & (1.8) & man\\
100737490 & 14.486 & 291.08209 & 0.69028 & 13.0 & 1.8 & man\\
100738231 & 14.643 & 291.08303 & 1.50924 & 27.6 & 3.3 & sap\\
100738670 & 14.847 & 291.08363 & 1.90112 & 29.7 & 3.7 & man\\
100742554 & 14.628 & 291.08879 & 1.76733 & 19.2 &  & man\\
100743629 & 14.043 & 291.09024 & 1.63491 & 27.6 &  & man\\
100745423 & 12.873 & 291.09263 & 1.75048 & 12.5 & 1.7 & man\\
100747150 & 14.378 & 291.09483 & 0.83749 & 18.4 & 2.7 & man\\
100752538 & 12.618 & 291.10197 & 1.89511 & 35.5 & (4.1) & sap\\
100758194 & 14.191 & 291.10964 & 1.70849 & 26.6 & 3.6 & sap\\
100758361 & 14.385 & 291.10991 & 1.40466 & 29.9 &  & man\\
100763478 & 13.192 & 291.11677 & 1.58127 & 32.4 & (3.7) & sap\\
100768145 & 13.163 & 291.12307 & 1.81926 & 38.6 & (4.4) & sap\\
100770682 & 14.266 & 291.12658 & 1.14876 & 26.7 & (2.7) & man\\
100773163 & 13.901 & 291.12987 & 1.10874 & 46.3 & 5.3 & sap\\
100782155 & 13.141 & 291.14165 & 1.11360 & 57.4 & 5.4 & sap\\
100784138 & 14.198 & 291.14421 & 1.68746 & 26.6 & 3.3 & man\\
100787298 & 14.936 & 291.14853 & 1.14182 & 28.4 & 3.7 & man\\
100790731 & 13.846 & 291.15304 & 1.17988 & 31.9 & (3.6) & sap\\
100790822 & 15.203 & 291.15315 & 1.47807 & 9.1 & (1.5) & man\\
100790832 & 14.226 & 291.15317 & 0.77703 & 57.4 & (5.7) & sap\\
100792637 & 12.883 & 291.15562 & 1.82894 & 52.9 & 5.1 & man\\
100793294 & 14.276 & 291.15646 & 0.71483 & 34.7 & 4.1 & sap\\
100795824 & 12.967 & 291.15980 & 1.89581 & 38.3 & 4.8 & man\\
100799833 & 12.183 & 291.16502 & 1.68434 & 33.0 & 3.8 & sap\\
100802610 & 14.893 & 291.16867 & 0.37907 & 29.9 &  & man\\
100805172 & 13.181 & 291.17204 & 1.57948 & 32.1 & 3.7 & sap\\
100807132 & 14.480 & 291.17456 & 1.84306 & 31.5 &  & man\\
100809477 & 15.081 & 291.17756 & 0.96286 & 31.3 & 3.8 & man\\
100809880 & 13.967 & 291.17808 & 1.99645 & 67.1 & 6.4 & man\\
100813027 & 15.011 & 291.18237 & 0.94779 & 25.0 & (3.3) & man\\
100813221 & 15.773 & 291.18264 & 1.69099 & 10.0 & 1.5 & man\\
100813661 & 14.298 & 291.18322 & 1.60183 & 27.9 & (3.6) & man\\
100813799 & 13.756 & 291.18340 & 1.35718 & 65.7 & 6.1 & man\\
100817208 & 14.083 & 291.18790 & 1.60311 & 48.2 & 5.4 & man\\
100819874 & 12.571 & 291.19141 & 1.59430 & 24.3 & 3.3 & sap\\
100820820 & 14.621 & 291.19264 & 1.22416 & 11.6 & 2.1 & man\\
100821572 & 13.969 & 291.19365 & 1.33072 & 53.2 &  & man\\
100824618 & 14.781 & 291.19754 & 1.51247 & 11.5 & 1.4 & man\\
100826011 & 15.096 & 291.19939 & 1.84490 & 50.1 &  & man\\
100826123 & 14.713 & 291.19953 & 1.46327 & 40.4 & (4.1) & man\\
100826589 & 13.366 & 291.20015 & 1.27822 & 27.7 & 3.5 & sap\\
100827073 & 14.331 & 291.20081 & 0.93753 & 34.0 & 3.8 & sap\\
100827490 & 13.453 & 291.20139 & 1.66865 & 15.0 & (2.0) & man\\
100828924 & 14.516 & 291.20327 & 1.15196 & 18.2 &  & man\\
100830101 & 14.552 & 291.20477 & 0.32313 & 26.5 & 3.0 & sap\\
100833997 & 13.806 & 291.20989 & 0.65472 & 31.4 & 3.9 & sap\\
100834084 & 12.548 & 291.21001 & 1.64546 & 31.8 & 4.0 & sap\\
100836428 & 14.242 & 291.21317 & 1.94311 & 31.6 & 4.0 & sap\\
100836619 & 14.229 & 291.21345 & 0.45605 & 28.0 & (3.7) & sap\\
100837771 & 13.696 & 291.21496 & 1.29365 & 38.8 & (4.2) & sap\\
100838216 & 14.361 & 291.21550 & 0.99303 & 37.8 & 4.1 & man\\
100838545 & 12.758 & 291.21593 & 1.49494 & 29.5 & (3.7) & sap\\
100841360 & 13.452 & 291.21953 & 1.95018 & 20.5 & 3.0 & sap\\
100841417 & 14.014 & 291.21961 & 0.43728 & 18.8 & 2.6 & man\\
100845030 & 14.456 & 291.22450 & 0.63361 & 30.6 & 3.9 & sap\\
100845233 & 13.688 & 291.22475 & 1.62023 & 28.5 & (3.0) & man\\
100846057 & 12.260 & 291.22584 & 1.54956 & 28.3 & (3.3) & sap\\
100849897 & 14.334 & 291.23092 & 0.46696 & 59.7 & 5.9 & sap\\
100852459 & 14.206 & 291.23427 & 1.58504 & 29.1 & (3.5) & man\\
100853452 & 12.629 & 291.23553 & 0.52005 & 32.3 &  & sap\\
100853745 & 13.779 & 291.23586 & 1.56867 & 39.7 & (4.6) & man\\
100855073 & 14.133 & 291.23768 & 1.51050 & 26.5 &  & sap\\
100856144 & 14.108 & 291.23909 & 1.36622 & 37.9 &  & man\\
100856234 & 13.359 & 291.23923 & 0.41364 & 102.4 & (8.9) & man\\
100856697 & 14.214 & 291.23982 & 0.87455 & 47.4 & 4.9 & man\\
100858245 & 13.742 & 291.24183 & 0.27329 & 31.5 & (3.2) & sap\\
100861153 & 14.596 & 291.24562 & 0.75779 & 36.9 & 4.1 & man\\
100861203 & 14.532 & 291.24569 & 0.21613 & 28.9 & (3.8) & man\\
100861335 & 13.042 & 291.24587 & 0.20847 & 34.8 & 4.1 & sap\\
100864569 & 13.254 & 291.25008 & 0.44017 & 63.7 &  & man\\
100867373 & 15.447 & 291.25370 & 1.91347 & 24.6 & (3.2) & man\\
100867895 & 13.712 & 291.25435 & 1.92338 & 33.9 & (3.9) & sap\\
100868399 & 14.032 & 291.25501 & 1.95535 & 14.3 & (2.2) & man\\
100871490 & 14.951 & 291.25914 & 1.13485 & 27.6 & (3.5) & man\\
100872561 & 15.537 & 291.26063 & 1.93630 & 12.7 & (1.9) & man\\
100873140 & 14.436 & 291.26135 & 0.75795 & 26.7 & 3.5 & man\\
100873731 & 13.641 & 291.26206 & 1.01897 & 35.2 & (3.8) & sap\\
100879189 & 14.548 & 291.26935 & 1.68646 & 23.1 & (3.3) & man\\
100880990 & 12.138 & 291.27175 & 1.61755 & 32.2 & 3.9 & sap\\
100885791 & 14.609 & 291.27815 & 1.33483 & 32.2 & 3.9 & sap\\
100886873 & 13.571 & 291.27959 & 1.10133 & 40.8 & 4.7 & sap\\
100886908 & 14.404 & 291.27965 & 0.42005 & 10.8 & 1.1 & man\\
100887322 & 13.337 & 291.28018 & 0.23459 & 40.2 &  & man\\
100887322 & 13.337 & 291.28018 & 0.23459 & 40.3 &  & sap\\
100888944 & 14.312 & 291.28229 & 0.23944 & 48.0 &  & man\\
100889852 & 14.630 & 291.28348 & 1.87466 & 53.0 & (5.6) & man\\
100892148 & 13.813 & 291.28663 & 1.78360 & 55.1 & 5.5 & man\\
100893246 & 14.806 & 291.28800 & 1.21694 & 32.2 & (3.8) & man\\
100893803 & 14.746 & 291.28877 & 1.85592 & 33.5 & 3.6 & man\\
100896955 & 14.831 & 291.29285 & 1.20868 & 26.3 &  & man\\
100898422 & 12.583 & 291.29480 & 1.67532 & 37.5 & 4.4 & sap\\
100899564 & 14.779 & 291.29634 & 1.54729 & 16.4 & (2.6) & man\\
100900153 & 13.616 & 291.29703 & 1.14535 & 27.2 &  & sap\\
100901855 & 14.174 & 291.29931 & 0.84220 & 25.2 & 3.3 & man\\
100901998 & 12.501 & 291.29949 & 0.94324 & 47.5 & 4.6 & sap\\
100902585 & 15.029 & 291.30027 & 0.41639 & 27.4 & (2.9) & man\\
100905864 & 14.037 & 291.30457 & 1.99818 & 30.0 & (3.9) & man\\
100908597 & 13.949 & 291.30820 & 0.56744 & 29.4 & 3.7 & sap\\
100911685 & 13.393 & 291.31240 & 2.01897 & 23.8 & 3.1 & man\\
100911815 & 12.959 & 291.31260 & 0.47401 & 22.6 & 3.3 & sap\\
100914315 & 14.419 & 291.31597 & 0.56954 & 35.4 &  & man\\
100914473 & 15.222 & 291.31617 & 0.21228 & 16.4 & 2.2 & man\\
100915882 & 14.766 & 291.31813 & 1.18749 & 29.5 & (3.5) & man\\
100920975 & 14.528 & 291.32494 & 1.77388 & 12.8 & (2.0) & man\\
100921071 & 13.993 & 291.32506 & 1.73479 & 29.3 & (3.7) & sap\\
100921343 & 12.914 & 291.32543 & 0.36513 & 30.5 & 3.8 & sap\\
100922068 & 14.447 & 291.32638 & 0.10787 & 26.5 & (3.4) & man\\
100922474 & 14.057 & 291.32692 & 0.28199 & 29.3 & (3.7) & man\\
100922926 & 12.896 & 291.32753 & 0.01955 & 31.0 & 3.5 & sap\\
100923801 & 14.074 & 291.32866 & 0.42331 & 32.8 &  & sap\\
100924078 & 13.267 & 291.32901 & 0.37681 & 17.4 & (2.4) & sap\\
100928979 & 13.007 & 291.33535 & 0.02051 & 35.7 & 4.3 & sap\\
100929178 & 14.710 & 291.33565 & 1.85446 & 12.0 & 1.7 & man\\
100929688 & 13.046 & 291.33638 & 1.32690 & 13.9 & 1.8 & man\\
100934222 & 14.121 & 291.34249 & 1.51435 & 36.7 & (3.9) & man\\
100935924 & 14.806 & 291.34490 & 1.33433 & 37.3 &  & man\\
100937183 & 13.632 & 291.34659 & 0.24906 & 34.0 & (4.0) & sap\\
100937360 & 14.283 & 291.34683 & 1.50686 & 25.1 & 3.6 & man\\
100940297 & 13.947 & 291.35072 & 0.15197 & 66.1 & 6.9 & man\\
100943381 & 13.383 & 291.35483 & 0.37929 & 15.6 & 1.9 & man\\
100945495 & 13.189 & 291.35774 & 0.45517 & 32.3 & (4.0) & sap\\
100948410 & 15.182 & 291.36163 & 1.98400 & 26.5 & (3.2) & man\\
100949965 & 14.200 & 291.36378 & 1.82220 & 11.0 & 1.5 & man\\
100953642 & 13.271 & 291.36866 & 0.87465 & 17.4 & (2.5) & sap\\
100954989 & 13.974 & 291.37053 & 1.50879 & 33.8 & (4.2) & sap\\
100956384 & 14.466 & 291.37237 & 0.80213 & 32.8 & (3.8) & sap\\
100958423 & 14.762 & 291.37519 & 0.35152 & 11.4 & 1.7 & man\\
100958571 & 14.660 & 291.37535 & 1.00666 & 14.5 & 2.2 & man\\
100958710 & 12.477 & 291.37553 & 0.08861 & 33.5 & (4.0) & sap\\
100959121 & 13.002 & 291.37607 & 2.02423 & 33.9 & 3.9 & sap\\
100960341 & 14.766 & 291.37768 & 1.27744 & 43.7 &  & man\\
100967582 & 14.060 & 291.38714 & 2.01422 & 20.7 & (2.9) & man\\
100968429 & 13.774 & 291.38827 & 1.35596 & 31.5 & (3.7) & sap\\
100968875 & 13.666 & 291.38887 & 1.32535 & 25.0 & 2.6 & man\\
100971166 & 14.262 & 291.39191 & 1.15215 & 15.2 & 2.0 & man\\
100971567 & 14.177 & 291.39248 & 1.77656 & 11.6 & (1.6) & man\\
100971774 & 13.651 & 291.39272 & 1.50388 & 52.2 & 5.4 & man\\
100973808 & 13.594 & 291.39547 & 1.43117 & 32.8 &  & sap\\
100974118 & 13.744 & 291.39585 & 0.36911 & 39.1 &  & sap\\
100980762 & 13.466 & 291.40460 & 1.96770 & 28.7 &  & man\\
100981445 & 14.328 & 291.40551 & 1.38470 & 30.1 & 3.9 & man\\
100983148 & 13.601 & 291.40783 & 0.92724 & 33.9 & 3.4 & sap\\
100983871 & 13.514 & 291.40879 & 1.69909 & 38.0 & 4.1 & sap\\
100985849 & 13.713 & 291.41140 & 1.94710 & 34.7 &  & sap\\
100986828 & 14.430 & 291.41277 & 1.79596 & 31.8 & (4.0) & man\\
100988199 & 14.296 & 291.41463 & 0.73395 & 28.1 & 3.6 & man\\
100988656 & 15.149 & 291.41524 & 0.51981 & 18.6 & (2.1) & man\\
100988725 & 13.274 & 291.41532 & 1.69215 & 26.1 & 3.4 & sap\\
100988784 & 14.415 & 291.41540 & 0.20058 & 58.5 & (5.8) & man\\
100988877 & 14.429 & 291.41553 & 1.24339 & 36.0 & (3.6) & sap\\
100989736 & 14.811 & 291.41675 & 0.95345 & 30.8 & (4.0) & man\\
100991403 & 14.502 & 291.41892 & 0.43622 & 26.5 & (3.0) & man\\
100991658 & 13.055 & 291.41925 & 1.38877 & 50.7 & 5.3 & sap\\
100992435 & 14.320 & 291.42033 & 1.78134 & 28.0 & (2.8) & man\\
100993976 & 13.380 & 291.42232 & 0.98060 & 22.4 & (3.2) & sap\\
100994184 & 15.579 & 291.42260 & 1.56377 & 13.1 & (1.3) & man\\
100997312 & 15.230 & 291.42682 & 0.94810 & 18.2 & 2.5 & man\\
100998571 & 13.704 & 291.42852 & -0.20099 & 23.5 & 2.9 & sap\\
101000321 & 12.780 & 291.43091 & 0.60447 & 39.4 & 4.9 & sap\\
101001054 & 12.848 & 291.43183 & 0.46790 & 38.9 & 4.3 & sap\\
101002050 & 13.274 & 291.43318 & -0.25271 & 24.9 & 3.2 & sap\\
101004170 & 14.752 & 291.43612 & -0.08080 & 42.8 & (4.3) & man\\
101005339 & 14.186 & 291.43780 & 1.89714 & 29.0 & (3.6) & man\\
101009305 & 13.590 & 291.44377 & 0.45081 & 9.7 & 1.5 & man\\
101009538 & 14.167 & 291.44420 & 1.46534 & 32.7 &  & man\\
101015137 & 14.840 & 291.45425 & 0.95231 & 44.8 & 4.7 & man\\
101015642 & 14.417 & 291.45515 & 0.15252 & 25.4 & (3.2) & man\\
101016219 & 14.007 & 291.45625 & 0.18111 & 36.3 & (4.0) & sap\\
101017465 & 14.092 & 291.45850 & 0.23151 & 9.0 & 1.3 & man\\
101017915 & 15.065 & 291.45932 & 2.12550 & 21.0 & (2.8) & man\\
101018056 & 15.042 & 291.45955 & 0.13176 & 24.4 & (3.1) & man\\
101018393 & 13.982 & 291.46019 & -0.02272 & 16.4 & 2.5 & man\\
101019013 & 13.906 & 291.46139 & -0.31073 & 41.1 & 4.3 & man\\
101020804 & 14.391 & 291.46472 & -0.18978 & 67.0 &  & man\\
101021652 & 13.677 & 291.46630 & 1.41317 & 28.6 & (3.7) & sap\\
101022193 & 13.245 & 291.46728 & 2.09455 & 34.6 & (4.0) & sap\\
101023768 & 13.942 & 291.47014 & 0.21349 & 41.4 & 4.6 & man\\
101023918 & 14.111 & 291.47042 & -0.26929 & 44.4 & 5.0 & man\\
101024376 & 12.797 & 291.47122 & 0.04472 & 18.2 & (2.6) & sap\\
101025128 & 14.415 & 291.47269 & 1.90547 & 32.9 & 3.4 & sap\\
101025264 & 14.572 & 291.47295 & 0.21938 & 20.4 & (2.2) & man\\
101025358 & 13.902 & 291.47314 & 0.24386 & 29.4 &  & sap\\
101025964 & 14.892 & 291.47426 & 0.28930 & 55.7 & (5.9) & man\\
101026664 & 14.486 & 291.47558 & 1.33582 & 51.8 & 5.0 & man\\
101027968 & 12.787 & 291.47797 & 0.53973 & 40.9 & 4.5 & sap\\
101029591 & 13.851 & 291.48099 & -0.27215 & 28.5 & (3.0) & man\\
101030924 & 14.397 & 291.48348 & 0.62523 & 69.3 & (6.2) & man\\
101032580 & 12.783 & 291.48659 & 1.11159 & 43.5 & 4.4 & sap\\
101033184 & 14.905 & 291.48775 & 1.91896 & 17.1 & 2.2 & man\\
101034439 & 13.955 & 291.49014 & 2.00285 & 29.5 & (3.8) & man\\
101034839 & 14.547 & 291.49086 & 0.19576 & 9.9 & 1.5 & man\\
101034881 & 13.500 & 291.49093 & 0.93401 & 43.5 & 4.7 & sap\\
101035230 & 13.628 & 291.49162 & 0.83570 & 46.9 & (5.1) & sap\\
101035304 & 13.195 & 291.49178 & 2.03472 & 24.0 &  & sap\\
101035766 & 12.906 & 291.49261 & 1.31397 & 10.2 & 1.5 & man\\
101037083 & 13.390 & 291.49512 & 1.59448 & 39.1 &  & sap\\
101039409 & 13.695 & 291.49944 & 2.06531 & 34.8 &  & sap\\
101039834 & 14.775 & 291.50017 & 2.07042 & 29.2 & 3.5 & man\\
101040751 & 15.037 & 291.50194 & 1.51269 & 11.5 & (1.1) & man\\
101041814 & 13.507 & 291.50399 & 1.52722 & 43.0 & 5.0 & sap\\
101042011 & 13.115 & 291.50433 & 2.04700 & 34.3 & (3.4) & sap\\
101043587 & 12.202 & 291.50727 & 0.55733 & 45.9 & (4.6) & man\\
101044584 & 12.996 & 291.50928 & -0.31709 & 25.2 & (3.3) & sap\\
101044694 & 15.106 & 291.50948 & 1.28669 & 18.1 & 2.1 & man\\
101044836 & 13.142 & 291.50975 & -0.08486 & 12.8 & (1.4) & man\\
101045095 & 13.967 & 291.51019 & 0.06251 & 21.3 & 2.8 & man\\
101046542 & 14.972 & 291.51296 & 0.55712 & 16.4 & (2.3) & man\\
101046557 & 13.265 & 291.51298 & 2.16556 & 61.8 & 6.5 & sap\\
101046788 & 14.392 & 291.51342 & 0.16831 & 22.1 & 3.2 & man\\
101047228 & 13.547 & 291.51423 & 1.40513 & 51.8 & 5.6 & sap\\
101050198 & 14.250 & 291.51984 & 1.56576 & 25.0 & (3.2) & man\\
101050222 & 13.281 & 291.51988 & -0.27829 & 47.9 & 4.7 & sap\\
101050462 & 14.495 & 291.52036 & 1.63532 & 28.5 &  & man\\
101050632 & 13.947 & 291.52068 & 0.47317 & 28.3 & 3.8 & sap\\
101051096 & 13.753 & 291.52161 & 1.70241 & 31.5 & (3.1) & sap\\
101054647 & 14.168 & 291.52828 & 0.87295 & 51.9 & 5.3 & sap\\
101056009 & 13.917 & 291.53086 & 1.44659 & 24.3 & (3.0) & sap\\
101056429 & 15.087 & 291.53170 & 0.61065 & 60.3 &  & man\\
101057165 & 14.067 & 291.53303 & 0.10175 & 18.4 & (2.5) & man\\
101057962 & 12.691 & 291.53458 & -0.42275 & 13.0 & (2.1) & man\\
101058180 & 12.613 & 291.53505 & 1.86233 & 31.2 & 3.9 & sap\\
101058880 & 13.680 & 291.53638 & 1.56013 & 43.2 & 4.3 & sap\\
101059151 & 13.353 & 291.53683 & 1.72685 & 26.7 & (3.5) & sap\\
101059381 & 12.622 & 291.53722 & -0.00047 & 30.1 & 3.9 & sap\\
101061858 & 13.995 & 291.54200 & 2.17327 & 25.5 & (3.1) & man\\
101062545 & 13.015 & 291.54330 & 2.15023 & 36.6 & 4.4 & sap\\
101063559 & 13.682 & 291.54524 & -0.03343 & 20.6 & (2.3) & sap\\
101064543 & 14.187 & 291.54719 & -0.10572 & 38.3 & 4.5 & sap\\
101064646 & 14.797 & 291.54739 & 1.52595 & 37.8 & (4.5) & man\\
101065131 & 14.497 & 291.54832 & 0.67604 & 24.9 & (2.9) & man\\
101065685 & 14.915 & 291.54940 & 1.55568 & 25.9 & 3.3 & man\\
101066347 & 13.847 & 291.55069 & 1.45060 & 55.9 & 5.7 & man\\
101067603 & 13.520 & 291.55316 & 0.84204 & 31.9 & 3.9 & sap\\
101069862 & 14.084 & 291.55770 & 1.82206 & 30.1 &  & sap\\
101071139 & 15.535 & 291.56076 & 2.12895 & 25.4 & (3.3) & man\\
101071751 & 13.951 & 291.56292 & -0.22354 & 13.4 & (1.5) & man\\
101072055 & 15.027 & 291.56414 & 1.47770 & 23.2 & (2.6) & man\\
101073282 & 14.010 & 291.56788 & 1.27910 & 56.6 & (5.6) & man\\
101073867 & 14.487 & 291.56904 & 0.67811 & 30.0 & 3.7 & sap\\
101077806 & 14.161 & 291.57535 & 1.33321 & 27.9 & (3.0) & sap\\
101078244 & 13.879 & 291.57602 & 2.08795 & 35.9 & 3.9 & sap\\
101079646 & 14.556 & 291.57824 & 1.84024 & 67.6 & 6.6 & man\\
101080756 & 13.850 & 291.58006 & 0.38919 & 55.5 & 6.0 & man\\
101081290 & 12.731 & 291.58092 & -0.54983 & 51.1 & 5.3 & sap\\
101084795 & 14.079 & 291.58641 & 0.58372 & 23.3 &  & man\\
101085186 & 14.753 & 291.58706 & 1.25963 & 34.0 & (3.7) & man\\
101087560 & 14.368 & 291.59076 & 0.04236 & 26.3 & (2.9) & man\\
101090302 & 15.298 & 291.59503 & -0.16770 & 31.5 & (3.8) & man\\
101090411 & 12.938 & 291.59520 & 1.85245 & 27.1 & (3.7) & sap\\
101090725 & 14.985 & 291.59567 & 0.88635 & 19.5 &  & man\\
101092562 & 14.238 & 291.59859 & 0.04762 & 16.9 & (2.3) & man\\
101092813 & 12.032 & 291.59895 & -0.05994 & 33.7 & (3.4) & sap\\
101093482 & 14.280 & 291.60003 & 1.60977 & 77.0 & 7.3 & man\\
101093867 & 14.150 & 291.60058 & 1.68520 & 27.5 &  & man\\
101094148 & 13.985 & 291.60098 & 2.02947 & 21.4 & 2.6 & sap\\
101098439 & 13.631 & 291.60761 & 1.04801 & 27.9 & (3.6) & sap\\
101098782 & 12.672 & 291.60819 & 0.24737 & 20.6 & (2.4) & sap\\
101099173 & 12.733 & 291.60879 & 1.95501 & 22.4 & 3.1 & man\\
101099720 & 12.906 & 291.60964 & -0.21804 & 34.2 & 4.1 & sap\\
101100065 & 13.274 & 291.61023 & 0.49958 & 34.9 & 4.1 & sap\\
101100415 & 14.877 & 291.61082 & 0.16362 & 27.1 & (2.7) & man\\
101101232 & 13.889 & 291.61213 & 2.05483 & 30.1 & (3.7) & sap\\
101101253 & 14.289 & 291.61217 & 1.54975 & 12.5 & (2.1) & man\\
101102567 & 14.031 & 291.61426 & -0.48679 & 15.7 & 2.2 & man\\
101103214 & 12.914 & 291.61532 & 2.01659 & 26.3 & 3.2 & sap\\
101106018 & 14.005 & 291.61961 & 2.19416 & 29.9 & (3.6) & sap\\
101106220 & 14.642 & 291.61992 & 0.21139 & 26.2 & (3.4) & man\\
101107208 & 13.082 & 291.62149 & 0.19295 & 32.3 & 4.0 & sap\\
101108197 & 13.826 & 291.62301 & -0.35573 & 17.8 & 2.7 & man\\
101108258 & 15.059 & 291.62312 & -0.62965 & 24.6 & (2.9) & man\\
101108472 & 15.037 & 291.62346 & 0.13823 & 34.3 & (3.4) & man\\
101111723 & 13.828 & 291.62856 & 1.06470 & 36.2 & (4.2) & man\\
101111963 & 13.605 & 291.62890 & 2.14874 & 17.5 & 2.5 & man\\
101112259 & 13.769 & 291.62938 & 1.57138 & 60.3 & 5.7 & man\\
101113062 & 13.462 & 291.63062 & -0.03739 & 72.3 & 7.2 & sap\\
101114620 & 14.725 & 291.63303 & 2.13688 & 28.3 & (3.8) & man\\
101115198 & 15.121 & 291.63398 & -0.39851 & 45.0 & 4.8 & man\\
101116845 & 14.038 & 291.63650 & 1.03435 & 37.7 & (4.5) & sap\\
101118040 & 14.152 & 291.63838 & 0.16601 & 26.5 &  & man\\
101119005 & 14.165 & 291.63994 & 0.59070 & 66.7 & 6.9 & man\\
101120312 & 14.939 & 291.64210 & 1.97412 & 27.4 & 3.5 & man\\
101121241 & 14.653 & 291.64349 & 2.01468 & 30.1 & 3.8 & man\\
101122421 & 12.824 & 291.64539 & 0.38612 & 43.1 & 4.7 & sap\\
101123395 & 12.939 & 291.64689 & 2.04155 & 46.3 & (4.7) & sap\\
101123432 & 13.488 & 291.64696 & 1.24289 & 26.2 &  & sap\\
101124344 & 13.176 & 291.64837 & -0.45699 & 37.6 & (4.2) & sap\\
101126093 & 15.401 & 291.65111 & -0.58210 & 15.4 & 2.2 & man\\
101126186 & 15.095 & 291.65127 & 1.56076 & 27.0 & 3.3 & man\\
101127623 & 13.652 & 291.65358 & 0.54981 & 35.1 & (3.5) & man\\
101130130 & 13.343 & 291.65749 & 1.25378 & 15.6 & (2.2) & man\\
101133310 & 14.504 & 291.66246 & 1.97811 & 39.6 & 4.3 & sap\\
101134441 & 14.500 & 291.66421 & 1.68784 & 39.0 & 4.4 & sap\\
101136306 & 12.611 & 291.66706 & 1.45305 & 30.8 & 4.0 & sap\\
101137245 & 14.115 & 291.66853 & 2.10456 & 27.4 & (3.3) & sap\\
101139228 & 12.559 & 291.67170 & -0.63344 & 30.2 & 3.6 & sap\\
101141206 & 14.242 & 291.67469 & 1.58847 & 22.5 & (3.0) & man\\
101141309 & 13.387 & 291.67488 & 0.29601 & 9.8 & 1.3 & man\\
101141846 & 14.017 & 291.67575 & 0.20753 & 29.5 &  & sap\\
101144124 & 14.375 & 291.67935 & 2.23243 & 27.9 &  & man\\
101144866 & 14.694 & 291.68057 & 0.32887 & 15.1 & 2.2 & man\\
101145517 & 14.108 & 291.68161 & 1.23084 & 29.6 &  & man\\
101147664 & 14.885 & 291.68493 & 1.64543 & 33.6 & (3.5) & man\\
101148382 & 13.473 & 291.68605 & 1.70738 & 34.6 & (4.2) & sap\\
101149674 & 14.756 & 291.68817 & 1.86692 & 24.6 & (3.3) & man\\
101150004 & 12.505 & 291.68868 & 2.22050 & 28.6 & 3.5 & sap\\
101150212 & 13.885 & 291.68899 & 1.77616 & 25.1 &  & man\\
101150411 & 14.240 & 291.68926 & 0.50928 & 26.3 & (3.0) & man\\
101150783 & 14.482 & 291.68987 & 1.85087 & 21.9 & 2.6 & man\\
101150795 & 14.108 & 291.68989 & 0.86599 & 25.2 & 3.2 & sap\\
101151242 & 14.018 & 291.69058 & 1.14597 & 42.8 & (4.2) & man\\
101151570 & 15.292 & 291.69110 & 0.28046 & 12.7 & 1.8 & man\\
101154362 & 13.131 & 291.69547 & -0.20668 & 29.4 & 3.6 & sap\\
101154828 & 13.622 & 291.69617 & 0.14189 & 16.8 & 2.3 & man\\
101162331 & 14.142 & 291.70800 & -0.09266 & 25.9 & (3.3) & man\\
101162392 & 12.718 & 291.70811 & 1.19743 & 45.6 & 4.8 & sap\\
101162838 & 14.798 & 291.70884 & 1.86259 & 23.9 &  & man\\
101163504 & 13.951 & 291.70989 & 0.02366 & 74.6 & 7.0 & man\\
101163959 & 13.971 & 291.71061 & 1.07502 & 28.7 & (3.5) & man\\
101164110 & 13.705 & 291.71085 & 2.19214 & 26.9 & (3.7) & man\\
101164298 & 14.457 & 291.71114 & 0.31052 & 26.6 &  & man\\
101165983 & 13.551 & 291.71389 & -0.19336 & 34.6 & 4.3 & sap\\
101166514 & 15.721 & 291.71474 & -0.29253 & 27.7 & 3.6 & man\\
101166955 & 14.778 & 291.71538 & 1.27191 & 10.5 & (1.0) & man\\
101167637 & 13.037 & 291.71646 & 0.73351 & 33.4 & 3.9 & man\\
101167976 & 15.455 & 291.71698 & 2.18577 & 33.5 & 3.9 & man\\
101169312 & 13.858 & 291.71908 & 0.88646 & 28.5 & (3.3) & man\\
101169337 & 14.554 & 291.71912 & 1.09439 & 53.7 & 5.5 & man\\
101169641 & 15.301 & 291.71956 & -0.55280 & 28.0 & 3.8 & man\\
101171937 & 14.205 & 291.72328 & 1.55262 & 33.4 &  & sap\\
101172152 & 14.127 & 291.72359 & 0.60772 & 30.4 & 3.7 & sap\\
101172950 & 14.204 & 291.72483 & 1.95543 & 27.1 & 3.3 & man\\
101173044 & 12.527 & 291.72497 & 0.19681 & 35.1 & (4.4) & sap\\
101174701 & 14.907 & 291.72757 & 0.28982 & 26.3 & 3.7 & man\\
101175503 & 13.660 & 291.72881 & 0.45476 & 26.2 & (3.1) & sap\\
101176093 & 13.139 & 291.72978 & 1.99585 & 22.5 & (2.9) & sap\\
101177353 & 14.854 & 291.73172 & 0.48137 & 33.8 & 3.8 & man\\
101178248 & 12.382 & 291.73310 & 0.10137 & 21.7 & 2.6 & sap\\
101179555 & 13.581 & 291.73508 & -0.39958 & 33.7 & 4.2 & man\\
101181202 & 14.437 & 291.73759 & -0.01584 & 22.9 & (3.0) & man\\
101181479 & 13.118 & 291.73807 & 0.04918 & 34.6 & 3.8 & sap\\
101184157 & 13.862 & 291.74223 & 0.16147 & 26.9 & (2.9) & sap\\
101187485 & 14.342 & 291.74754 & 0.59697 & 27.8 & 3.6 & sap\\
101187777 & 13.860 & 291.74796 & 1.62582 & 31.4 & (4.0) & sap\\
101188939 & 14.715 & 291.74977 & 1.67190 & 30.7 & (3.3) & sap\\
101189140 & 14.964 & 291.75011 & 0.35472 & 28.1 & (3.0) & man\\
101189811 & 14.071 & 291.75112 & -0.42451 & 34.1 & (4.3) & man\\
101191330 & 14.452 & 291.75344 & 0.72945 & 30.2 & (3.4) & man\\
101192102 & 13.978 & 291.75470 & 1.11048 & 30.6 & (3.5) & sap\\
101192695 & 13.409 & 291.75556 & 2.05615 & 33.7 & (4.2) & sap\\
101193334 & 14.642 & 291.75656 & 0.58980 & 53.1 & 5.7 & man\\
101194293 & 14.732 & 291.75803 & 1.83780 & 26.8 & 3.5 & man\\
101194709 & 14.719 & 291.75870 & 2.06235 & 23.2 & (2.7) & man\\
101195199 & 13.029 & 291.75951 & 1.97382 & 28.6 &  & sap\\
101195523 & 13.558 & 291.76002 & 1.83068 & 37.1 & 4.3 & sap\\
101196210 & 14.672 & 291.76113 & 0.66281 & 29.8 &  & man\\
101197556 & 13.351 & 291.76320 & -0.54024 & 34.1 & 3.8 & sap\\
101197732 & 13.661 & 291.76348 & -0.51246 & 15.9 & (2.1) & man\\
101199197 & 14.202 & 291.76571 & 0.60905 & 28.8 &  & man\\
101200044 & 14.406 & 291.76701 & 1.43179 & 28.6 & (3.4) & man\\
101201987 & 14.311 & 291.77004 & -0.11180 & 15.1 & (2.2) & man\\
101202518 & 14.944 & 291.77087 & 1.89952 & 28.6 & (3.3) & man\\
101203233 & 14.285 & 291.77201 & 1.75953 & 58.3 & (5.5) & sap\\
101204408 & 13.579 & 291.77378 & 0.53474 & 17.5 & 2.3 & man\\
101207562 & 12.718 & 291.77866 & 1.28381 & 32.7 & (4.1) & sap\\
101207636 & 13.951 & 291.77877 & -0.09446 & 49.9 & 5.4 & man\\
101208801 & 14.203 & 291.78065 & 1.14579 & 14.1 & (2.3) & man\\
101212236 & 13.905 & 291.78602 & 1.86924 & 34.9 & 4.0 & sap\\
101214403 & 13.788 & 291.78940 & 1.10562 & 21.0 & (2.8) & man\\
101214882 & 14.832 & 291.79015 & 0.21392 & 36.4 & (4.1) & man\\
101216430 & 14.988 & 291.79263 & 1.37272 & 25.0 &  & man\\
101217274 & 13.932 & 291.79390 & 0.59959 & 88.8 & (8.1) & man\\
101218811 & 13.462 & 291.79639 & 1.86938 & 34.0 & 3.9 & sap\\
101219553 & 13.894 & 291.79762 & 0.44281 & 102.3 & (9.1) & man\\
101222229 & 14.746 & 291.80178 & -0.45017 & 34.0 & (4.0) & man\\
101222796 & 13.065 & 291.80264 & 0.42683 & 12.9 & 1.5 & man\\
101223476 & 14.946 & 291.80364 & -0.19615 & 16.8 & 2.2 & man\\
101226184 & 15.033 & 291.80793 & 1.31134 & 27.1 & (3.4) & man\\
101229068 & 13.501 & 291.81262 & -0.54603 & 51.4 & 5.3 & sap\\
101229072 & 13.428 & 291.81263 & 1.19002 & 74.3 & 6.9 & man\\
101229714 & 13.822 & 291.81360 & 0.15832 & 27.8 & 3.5 & sap\\
101231842 & 12.473 & 291.81682 & 1.97938 & 30.8 & 3.5 & sap\\
101232297 & 12.476 & 291.81754 & -0.26094 & 37.7 &  & sap\\
101232883 & 14.527 & 291.81846 & 1.45619 & 28.7 &  & man\\
101233006 & 13.837 & 291.81862 & -0.07616 & 23.8 & (3.0) & man\\
101234832 & 13.251 & 291.82150 & -0.46851 & 30.7 & 3.8 & sap\\
101235724 & 13.378 & 291.82294 & 1.85331 & 36.3 & 4.2 & sap\\
101237841 & 12.706 & 291.82626 & -0.31012 & 13.7 & 1.5 & sap\\
101238328 & 12.627 & 291.82705 & 0.15114 & 37.2 & (3.8) & sap\\
101239347 & 14.198 & 291.82864 & 1.19585 & 27.6 & 3.5 & man\\
101239581 & 14.277 & 291.82904 & 0.75192 & 26.3 &  & man\\
101241291 & 14.592 & 291.83182 & 1.88469 & 25.0 &  & man\\
101242228 & 13.097 & 291.83330 & 0.75421 & 28.8 & (3.7) & sap\\
101243695 & 13.623 & 291.83566 & 1.25006 & 13.7 & 1.8 & man\\
101245682 & 14.479 & 291.83889 & 1.95751 & 22.3 & (2.4) & man\\
101245919 & 13.657 & 291.83930 & 0.70098 & 33.7 & 4.1 & sap\\
101246426 & 14.245 & 291.84015 & 0.49469 & 27.0 & 3.4 & sap\\
101246686 & 14.351 & 291.84056 & -0.53971 & 28.7 &  & man\\
101246687 & 14.865 & 291.84056 & 1.53945 & 30.0 & (3.1) & man\\
101247114 & 14.167 & 291.84118 & -0.08511 & 32.0 & 4.0 & man\\
101248026 & 13.281 & 291.84260 & -0.51509 & 50.8 & (5.0) & sap\\
101248294 & 13.559 & 291.84304 & 1.89417 & 30.3 & (3.8) & sap\\
101251252 & 13.721 & 291.84776 & 1.32805 & 64.6 & 6.5 & sap\\
101252836 & 13.833 & 291.85026 & 0.99847 & 22.7 & (3.1) & sap\\
101253078 & 14.121 & 291.85064 & -0.51641 & 15.4 & 2.0 & man\\
101253113 & 14.526 & 291.85068 & -0.41695 & 27.1 &  & man\\
101254020 & 13.576 & 291.85211 & -0.46991 & 43.0 & 4.7 & sap\\
101258619 & 12.757 & 291.85924 & -0.00033 & 37.7 & 4.4 & sap\\
101260646 & 13.588 & 291.86251 & 1.36033 & 87.7 & 8.1 & man\\
101262546 & 14.217 & 291.86558 & 0.73952 & 21.2 & (2.8) & man\\
101262678 & 13.761 & 291.86579 & -0.07394 & 34.4 & (4.1) & sap\\
101262795 & 14.156 & 291.86599 & 1.00213 & 30.4 & 3.8 & sap\\
101264567 & 14.947 & 291.86885 & 0.40868 & 30.1 & 4.0 & man\\
101265048 & 12.997 & 291.86964 & 0.27481 & 33.1 & 4.1 & sap\\
101265141 & 13.017 & 291.86979 & 0.29741 & 24.6 & 3.0 & sap\\
101267794 & 15.175 & 291.87398 & 0.51590 & 23.6 &  & man\\
101270424 & 13.143 & 291.87795 & 1.68065 & 33.7 & 4.1 & sap\\
101270438 & 12.889 & 291.87798 & 0.03079 & 15.1 & 2.0 & sap\\
101271394 & 13.488 & 291.87939 & -0.06955 & 24.8 & (3.1) & sap\\
101272355 & 15.081 & 291.88072 & 0.82882 & 25.1 & 3.1 & man\\
101273102 & 12.581 & 291.88183 & -0.47210 & 26.9 & (3.3) & sap\\
101273107 & 14.206 & 291.88184 & -0.31577 & 9.5 & (1.0) & man\\
101275308 & 12.828 & 291.88504 & 1.40820 & 25.4 & (3.3) & sap\\
101275997 & 13.529 & 291.88602 & 0.73522 & 20.1 & (2.5) & sap\\
101278229 & 14.591 & 291.88925 & 0.56474 & 25.9 & 3.1 & man\\
101279871 & 15.366 & 291.89169 & 1.17678 & 16.7 & 2.4 & man\\
101279963 & 13.331 & 291.89183 & -0.52641 & 52.1 & (5.2) & sap\\
101281195 & 14.370 & 291.89367 & 0.39727 & 15.9 & 2.2 & man\\
101281824 & 12.800 & 291.89463 & 1.32962 & 27.4 & 3.3 & sap\\
101281996 & 13.705 & 291.89490 & -0.03212 & 28.7 & (3.2) & sap\\
101286739 & 13.240 & 291.90217 & 0.95468 & 30.5 & (3.8) & sap\\
101287608 & 15.664 & 291.90341 & 0.20226 & 13.0 & 2.0 & man\\
101289231 & 13.782 & 291.90589 & 1.58396 & 28.9 & (3.8) & man\\
101289267 & 13.401 & 291.90596 & 1.71057 & 35.6 & (4.2) & sap\\
101289675 & 13.013 & 291.90657 & 0.94287 & 27.6 &  & sap\\
101290029 & 13.174 & 291.90710 & 1.52946 & 27.6 & (3.6) & sap\\
101290292 & 13.704 & 291.90751 & 0.59387 & 19.9 & 2.5 & man\\
101290847 & 13.204 & 291.90835 & 1.21587 & 60.1 & 5.8 & sap\\
101291471 & 12.502 & 291.90929 & 1.03558 & 36.8 & (4.3) & sap\\
101292808 & 14.639 & 291.91129 & 1.01110 & 31.0 & (3.9) & man\\
101295016 & 12.842 & 291.91455 & -0.02784 & 24.6 & (3.0) & sap\\
101296374 & 14.818 & 291.91662 & 0.46522 & 21.7 & (2.9) & man\\
101299401 & 13.194 & 291.92123 & 0.13697 & 26.7 & (3.7) & sap\\
101300965 & 14.804 & 291.92364 & 0.05322 & 25.3 & (3.4) & man\\
101301023 & 14.178 & 291.92374 & -0.02262 & 34.1 &  & man\\
101303810 & 15.400 & 291.92815 & 1.54780 & 26.2 &  & man\\
101303931 & 14.845 & 291.92835 & 1.25212 & 28.8 & 3.9 & man\\
101304467 & 14.966 & 291.92923 & 0.72475 & 32.6 & (4.1) & man\\
101304738 & 14.515 & 291.92964 & -0.12426 & 28.2 &  & man\\
101305294 & 13.725 & 291.93049 & 1.76328 & 36.3 & (4.1) & sap\\
101305676 & 12.306 & 291.93106 & 0.21048 & 30.0 & 3.9 & sap\\
101308337 & 14.296 & 291.93533 & -0.31556 & 26.8 & (2.8) & sap\\
101309464 & 13.939 & 291.93720 & 0.17960 & 33.6 & (4.3) & man\\
101312180 & 13.927 & 291.94150 & 0.31710 & 34.6 & (4.1) & man\\
101312308 & 14.214 & 291.94169 & -0.41716 & 17.1 & (2.1) & man\\
101312396 & 13.708 & 291.94182 & 1.41227 & 34.6 & 4.2 & sap\\
101312732 & 14.625 & 291.94236 & 0.18405 & 18.8 & 2.7 & man\\
101314527 & 13.392 & 291.94555 & -0.07234 & 59.2 &  & man\\
101315033 & 14.529 & 291.94652 & 1.42221 & 25.2 & (2.8) & man\\
101315056 & 13.908 & 291.94658 & 1.27239 & 26.7 & 3.1 & sap\\
101316136 & 13.301 & 291.94885 & 0.06353 & 28.2 & (2.9) & man\\
101316534 & 13.995 & 291.94970 & 1.76776 & 20.4 & (2.5) & man\\
101316698 & 13.418 & 291.95001 & 1.26016 & 32.6 & 4.1 & sap\\
101319931 & 14.705 & 291.95660 & 0.72060 & 30.1 & (4.0) & man\\
101319975 & 13.411 & 291.95672 & -0.42760 & 30.6 &  & sap\\
101320312 & 13.108 & 291.95742 & 1.32746 & 27.6 & 3.8 & sap\\
101320378 & 13.251 & 291.95756 & 0.32240 & 32.9 & 4.1 & man\\
101320586 & 13.354 & 291.95804 & 1.35960 & 58.3 & (6.2) & sap\\
101321936 & 14.252 & 291.96075 & 1.55702 & 28.5 & (3.7) & man\\
101322703 & 12.815 & 291.96227 & 1.05358 & 45.2 & 4.6 & sap\\
101322820 & 13.976 & 291.96251 & 0.18840 & 31.2 & (3.3) & sap\\
101324332 & 12.771 & 291.96545 & 0.26763 & 52.4 & (5.3) & sap\\
101325039 & 14.986 & 291.96688 & 0.69927 & 17.1 & (1.8) & man\\
101326028 & 14.331 & 291.96885 & -0.37445 & 33.6 & (4.0) & man\\
101326609 & 13.931 & 291.97006 & -0.36128 & 23.6 & (3.0) & sap\\
101326879 & 13.241 & 291.97062 & 0.60193 & 34.4 & 4.1 & sap\\
101331140 & 14.691 & 291.97935 & -0.21837 & 56.3 & (5.9) & man\\
101332107 & 12.884 & 291.98126 & 0.50582 & 29.5 &  & sap\\
101332727 & 15.728 & 291.98246 & 0.64177 & 30.3 & 3.7 & man\\
101332883 & 13.420 & 291.98278 & 0.72298 & 41.0 & 4.5 & sap\\
101335441 & 13.481 & 291.98791 & -0.24230 & 55.2 & 5.9 & sap\\
101335883 & 14.860 & 291.98879 & 1.12626 & 52.0 & 5.1 & man\\
101336091 & 12.891 & 291.98920 & -0.19441 & 35.3 & (4.0) & sap\\
101336195 & 13.772 & 291.98940 & 0.01402 & 17.6 & (2.3) & man\\
101339201 & 12.448 & 291.99554 & 1.39588 & 44.7 & 4.8 & sap\\
101339420 & 13.281 & 291.99600 & -0.44769 & 17.1 & (2.3) & man\\
101340626 & 14.056 & 291.99855 & 0.75001 & 23.3 & 2.7 & man\\
101343519 & 13.655 & 292.00464 & 1.09325 & 30.2 &  & sap\\
101347468 & 13.481 & 292.01276 & -0.25095 & 34.0 & 4.3 & sap\\
101347642 & 12.721 & 292.01310 & -0.35930 & 31.9 & 3.9 & sap\\
101347760 & 13.031 & 292.01333 & 0.26870 & 27.1 & (2.9) & sap\\
101349437 & 14.117 & 292.01666 & 1.61570 & 21.1 & 3.1 & man\\
101352174 & 13.196 & 292.02216 & -0.35694 & 31.1 & 3.9 & sap\\
101353737 & 12.952 & 292.02529 & 1.02912 & 62.3 & 5.8 & sap\\
101355793 & 14.674 & 292.02954 & 1.33194 & 26.0 &  & man\\
101356220 & 13.861 & 292.03043 & 0.69364 & 23.5 & (2.4) & man\\
101356616 & 14.121 & 292.03123 & 0.88493 & 26.4 & (3.5) & man\\
101358149 & 14.352 & 292.03431 & -0.07450 & 32.1 & (3.8) & man\\
101362519 & 14.420 & 292.04321 & 0.61949 & 28.4 & (3.8) & man\\
101362522 & 13.898 & 292.04322 & 0.83149 & 49.9 & 5.5 & sap\\
101363981 & 14.580 & 292.04619 & 1.04002 & 17.6 & 1.8 & man\\
101364068 & 13.521 & 292.04636 & -0.33343 & 21.8 & (2.8) & sap\\
101365233 & 14.592 & 292.04878 & 0.90574 & 15.6 & 2.4 & man\\
101365347 & 15.231 & 292.04900 & -0.20824 & 20.8 & (2.9) & man\\
101366598 & 13.849 & 292.05161 & 1.48692 & 32.3 & (3.9) & sap\\
101368062 & 14.511 & 292.05493 & -0.33227 & 27.3 & (2.8) & man\\
101368866 & 14.851 & 292.05684 & 0.25502 & 29.5 & (3.4) & man\\
101368951 & 12.407 & 292.05703 & 0.45265 & 35.4 &  & sap\\
101369568 & 14.521 & 292.05857 & -0.38716 & 31.0 &  & man\\
101369666 & 14.290 & 292.05885 & 0.58117 & 15.6 & (2.4) & man\\
101371690 & 13.095 & 292.06471 & 0.42102 & 33.0 & (3.5) & sap\\
101372310 & 13.456 & 292.06654 & 0.76842 & 44.0 & 4.6 & sap\\
101372675 & 15.062 & 292.06745 & 0.09699 & 33.7 & 4.0 & man\\
101373582 & 15.071 & 292.06940 & 0.38306 & 27.7 & 3.0 & man\\
101376555 & 13.378 & 292.07480 & 0.82183 & 32.4 & 4.0 & sap\\
101378314 & 14.105 & 292.07787 & 1.18485 & 17.8 & (2.0) & man\\
101378387 & 13.551 & 292.07798 & -0.37571 & 33.3 & (4.0) & sap\\
101378749 & 13.745 & 292.07864 & -0.41542 & 22.8 & 2.9 & sap\\
101378905 & 13.013 & 292.07888 & 1.37995 & 26.8 & 2.9 & sap\\
101378942 & 12.755 & 292.07893 & 1.09052 & 13.9 & 2.0 & sap\\
101378942 & 12.755 & 292.07893 & 1.09052 & 13.9 & 2.0 & man\\
101380751 & 13.271 & 292.08202 & 0.02348 & 42.1 &  & sap\\
101381379 & 14.900 & 292.08309 & 1.04161 & 22.8 & 3.0 & man\\
101385073 & 14.085 & 292.08946 & 0.64575 & 38.4 & 4.3 & sap\\
101385320 & 14.728 & 292.08990 & 0.60888 & 19.6 & 2.7 & man\\
101385870 & 14.495 & 292.09083 & 0.87539 & 30.2 & (3.0) & man\\
101386354 & 12.706 & 292.09167 & -0.36107 & 20.5 &  & sap\\
101387138 & 14.558 & 292.09297 & 1.46212 & 27.3 & (3.6) & man\\
101391721 & 14.097 & 292.10092 & -0.00401 & 25.6 &  & sap\\
101392868 & 13.866 & 292.10289 & 0.14535 & 23.5 & 2.6 & sap\\
101393792 & 14.571 & 292.10450 & 0.43682 & 24.7 & (2.5) & man\\
101398481 & 12.452 & 292.11263 & 0.89308 & 35.6 & (3.7) & sap\\
101402017 & 12.927 & 292.11881 & 0.00253 & 49.6 & (5.3) & sap\\
101402532 & 13.563 & 292.11970 & 1.35305 & 27.7 &  & man\\
101403032 & 15.216 & 292.12058 & 0.12008 & 31.5 & 3.8 & man\\
101403073 & 14.537 & 292.12066 & 0.89040 & 33.9 & (4.1) & man\\
101404415 & 15.371 & 292.12298 & -0.28002 & 14.2 & 2.1 & man\\
101405957 & 13.794 & 292.12572 & 0.52333 & 38.5 & 4.6 & man\\
101406318 & 12.601 & 292.12635 & -0.05400 & 35.7 & 4.0 & sap\\
101407436 & 13.896 & 292.12823 & -0.35164 & 25.6 & 3.3 & sap\\
101408422 & 13.622 & 292.12989 & -0.04569 & 56.6 & 5.7 & man\\
101409181 & 13.051 & 292.13122 & -0.14648 & 68.0 & 6.6 & sap\\
101411168 & 12.447 & 292.13457 & 0.14694 & 34.3 & (4.1) & sap\\
101411659 & 12.511 & 292.13531 & -0.33381 & 61.8 & (5.6) & man\\
101412091 & 14.064 & 292.13609 & 0.93705 & 28.5 &  & man\\
101413056 & 14.902 & 292.13770 & -0.01829 & 12.1 & 1.6 & man\\
101415641 & 13.852 & 292.14207 & 0.70118 & 29.9 & 3.8 & sap\\
101417670 & 13.811 & 292.14555 & 0.70250 & 13.2 & 1.9 & man\\
101418465 & 13.171 & 292.14694 & -0.26049 & 59.3 & 5.9 & sap\\
101419045 & 14.261 & 292.14794 & 0.42470 & 29.1 & (3.7) & man\\
101422280 & 15.116 & 292.15342 & -0.21793 & 67.2 & (6.7) & man\\
101423629 & 13.392 & 292.15564 & 0.14282 & 30.4 & (3.9) & sap\\
101424698 & 14.149 & 292.15755 & 0.52573 & 74.4 & 7.4 & man\\
101425849 & 14.878 & 292.15957 & 0.32844 & 22.4 & 3.0 & man\\
101426117 & 14.581 & 292.16008 & 0.11045 & 17.2 & 2.7 & man\\
101426170 & 13.236 & 292.16017 & 0.34969 & 35.1 & 4.0 & sap\\
101427312 & 14.315 & 292.16209 & -0.10034 & 30.3 & 3.6 & man\\
101427327 & 13.900 & 292.16211 & 1.30295 & 16.9 & 2.4 & man\\
101427969 & 14.876 & 292.16329 & 0.39829 & 63.4 & 6.2 & man\\
101430573 & 12.578 & 292.16786 & -0.17203 & 22.0 & (2.6) & sap\\
101432067 & 13.667 & 292.17052 & 0.19722 & 26.0 & (3.2) & man\\
101433432 & 13.521 & 292.17275 & 0.04984 & 31.2 & (3.7) & sap\\
101433941 & 13.710 & 292.17354 & 1.07809 & 29.2 &  & man\\
101439884 & 13.781 & 292.18389 & -0.31834 & 33.6 & 3.9 & sap\\
101440659 & 13.521 & 292.18521 & 0.92757 & 8.9 & 1.4 & man\\
101441726 & 14.722 & 292.18700 & -0.06462 & 27.5 &  & man\\
101442365 & 12.503 & 292.18808 & 0.63520 & 38.9 & 4.3 & sap\\
101442374 & 14.161 & 292.18810 & -0.17184 & 25.5 & 2.7 & man\\
101444217 & 14.541 & 292.19130 & 0.47664 & 51.6 &  & man\\
101446191 & 13.718 & 292.19465 & 0.61590 & 31.1 & 4.0 & sap\\
101446216 & 14.181 & 292.19470 & 0.71501 & 24.7 & (3.4) & man\\
101447328 & 14.641 & 292.19653 & 0.54924 & 25.7 &  & man\\
101448392 & 14.601 & 292.19826 & -0.36813 & 31.0 & 3.9 & man\\
101448417 & 13.947 & 292.19830 & 0.40492 & 53.6 & 5.7 & sap\\
101451115 & 13.395 & 292.20289 & 0.08869 & 25.6 & (3.4) & man\\
101451373 & 12.347 & 292.20331 & 0.49616 & 32.4 & (4.1) & sap\\
101451533 & 13.837 & 292.20357 & 0.46766 & 40.2 &  & sap\\
101454067 & 13.631 & 292.20777 & 0.37357 & 32.3 & (4.1) & sap\\
101458937 & 13.222 & 292.21625 & 0.17364 & 14.5 & 2.1 & man\\
101460486 & 13.503 & 292.21880 & 0.81992 & 34.7 & 4.0 & sap\\
101460775 & 14.761 & 292.21929 & 0.51695 & 42.3 & 4.6 & man\\
101460879 & 14.585 & 292.21943 & 0.83847 & 65.4 & 6.5 & man\\
101465171 & 13.802 & 292.22675 & 0.09967 & 21.5 &  & man\\
101467619 & 14.261 & 292.23090 & 0.04907 & 31.7 & 4.0 & man\\
101469820 & 14.360 & 292.23460 & 1.11314 & 16.5 & 2.1 & man\\
101473616 & 13.945 & 292.24089 & 0.52904 & 18.7 & 2.3 & man\\
101474112 & 13.892 & 292.24176 & 0.10558 & 62.4 & 6.5 & man\\
101474730 & 14.151 & 292.24286 & 0.24934 & 22.6 & 2.6 & man\\
101476440 & 12.951 & 292.24578 & 0.46899 & 12.3 & 1.7 & sap\\
101476920 & 14.838 & 292.24660 & 0.61513 & 15.3 & 2.1 & man\\
101477247 & 14.987 & 292.24719 & 0.71147 & 14.9 & (1.6) & man\\
101478540 & 12.721 & 292.24933 & -0.32656 & 22.7 & 2.7 & sap\\
101479332 & 13.851 & 292.25070 & 0.54807 & 32.7 & 4.1 & sap\\
101479386 & 13.007 & 292.25080 & 0.46425 & 26.7 & (3.4) & sap\\
101479567 & 12.611 & 292.25106 & 0.74391 & 25.1 & (3.1) & man\\
101479905 & 14.088 & 292.25164 & 0.97540 & 26.3 &  & man\\
101480480 & 14.991 & 292.25263 & 0.28173 & 16.3 & 2.2 & man\\
101480733 & 14.071 & 292.25310 & 0.59013 & 44.7 & 5.0 & man\\
101481433 & 13.846 & 292.25433 & -0.18910 & 32.7 & (3.9) & sap\\
101483826 & 12.707 & 292.25852 & 0.41784 & 26.9 & (3.5) & sap\\
101490360 & 12.744 & 292.26966 & 0.91643 & 59.7 & 6.0 & sap\\
101495773 & 13.786 & 292.27900 & -0.07316 & 38.1 & 4.1 & sap\\
101496643 & 13.315 & 292.28042 & 0.02775 & 25.6 & 3.3 & sap\\
101499895 & 13.527 & 292.28593 & 0.19790 & 64.9 & 6.1 & sap\\
101503482 & 13.753 & 292.29207 & 0.59820 & 22.8 & 3.0 & sap\\
101509360 & 13.056 & 292.30230 & -0.18344 & 55.5 & 5.7 & sap\\
101511309 & 14.298 & 292.30568 & 0.66261 & 35.4 & 4.1 & sap\\
101513155 & 12.978 & 292.30891 & 0.97732 & 32.5 & 4.2 & sap\\
101513442 & 14.316 & 292.30940 & 0.16418 & 30.0 &  & sap\\
101520144 & 12.801 & 292.32107 & 0.83685 & 13.3 & (1.9) & man\\
101521149 & 14.755 & 292.32290 & 0.60598 & 32.0 & (3.8) & man\\
101522290 & 12.122 & 292.32492 & -0.03924 & 24.9 &  & sap\\
101523962 & 12.267 & 292.32779 & 0.92039 & 27.0 & (2.9) & sap\\
101525862 & 14.023 & 292.33117 & 0.81941 & 19.7 &  & man\\
101528536 & 15.171 & 292.33599 & 0.51945 & 24.6 &  & man\\
101529924 & 14.206 & 292.33835 & -0.29341 & 24.6 &  & man\\
101536163 & 14.471 & 292.34943 & 0.01349 & 21.4 & 2.8 & man\\
101536782 & 14.781 & 292.35050 & 0.15965 & 22.3 & 2.6 & man\\
101538074 & 13.358 & 292.35279 & -0.16142 & 30.5 & (3.8) & sap\\
101538346 & 14.181 & 292.35333 & 0.28585 & 28.1 & 3.3 & man\\
101538547 & 14.217 & 292.35370 & -0.01102 & 10.6 & 1.7 & man\\
101538672 & 13.831 & 292.35390 & -0.18196 & 35.6 & 3.7 & sap\\
101539993 & 14.751 & 292.35622 & -0.22621 & 23.4 &  & man\\
101542075 & 14.827 & 292.35992 & 0.49076 & 12.8 & 1.9 & man\\
101544311 & 13.749 & 292.36378 & 0.56235 & 33.2 & 4.0 & sap\\
101546354 & 13.351 & 292.36739 & 0.74518 & 29.2 & (3.6) & sap\\
101546964 & 12.901 & 292.36851 & 0.61542 & 24.7 & 2.8 & sap\\
101550759 & 13.134 & 292.37523 & 0.07752 & 23.1 & (3.2) & sap\\
101554715 & 14.398 & 292.38161 & 0.54842 & 29.4 &  & man\\
101557699 & 15.160 & 292.38631 & 0.70108 & 33.0 & (4.1) & man\\
101557896 & 13.091 & 292.38659 & -0.21769 & 35.2 & (3.7) & sap\\
101558507 & 13.805 & 292.38756 & 0.74707 & 34.0 & 4.2 & sap\\
101561050 & 12.468 & 292.39166 & 0.25915 & 32.8 & 4.0 & sap\\
101561081 & 14.313 & 292.39170 & 0.09368 & 32.4 & (3.8) & man\\
101562508 & 13.141 & 292.39400 & 0.14149 & 26.9 & 3.7 & sap\\
101563951 & 13.816 & 292.39623 & 0.64370 & 28.5 & 3.6 & sap\\
101565025 & 13.626 & 292.39802 & -0.29366 & 25.8 &  & sap\\
101566235 & 14.663 & 292.39992 & 0.72789 & 30.3 & (4.0) & man\\
101568144 & 14.259 & 292.40298 & 0.75055 & 21.2 & 3.2 & man\\
101569001 & 14.191 & 292.40440 & 0.60189 & 38.3 & (4.0) & sap\\
101569925 & 13.162 & 292.40580 & 0.51492 & 32.4 & (3.2) & sap\\
101575148 & 13.421 & 292.41431 & -0.17525 & 28.5 & (3.0) & sap\\
101579183 & 13.535 & 292.42087 & 0.08014 & 27.8 & (3.5) & sap\\
101579756 & 13.394 & 292.42182 & -0.22418 & 28.4 & 3.7 & sap\\
101587725 & 14.736 & 292.43463 & 0.43942 & 21.5 & 2.7 & man\\
101589817 & 13.574 & 292.43818 & 0.12807 & 32.7 & (3.2) & sap\\
101593684 & 14.586 & 292.44514 & -0.01746 & 32.6 & 4.1 & man\\
101594911 & 13.688 & 292.44738 & 0.64048 & 31.1 & 3.7 & sap\\
101601779 & 14.161 & 292.46087 & 0.16955 & 14.9 & 2.1 & man\\
101602667 & 14.651 & 292.46271 & -0.17588 & 26.0 & 3.5 & man\\
101602989 & 14.646 & 292.46334 & 0.10876 & 59.3 & (6.2) & man\\
101606835 & 14.431 & 292.47106 & -0.20674 & 27.9 &  & man\\
101610551 & 12.301 & 292.47868 & -0.03471 & 17.6 & 2.1 & man\\
101611062 & 14.597 & 292.47972 & 0.35666 & 32.5 & (4.1) & man\\
101612565 & 13.132 & 292.48275 & 0.39426 & 22.2 & 2.9 & sap\\
101614830 & 13.861 & 292.48737 & 0.20515 & 31.6 & (3.5) & man\\
101615168 & 14.571 & 292.48798 & -0.25639 & 22.3 & 3.0 & man\\
101615645 & 13.242 & 292.48895 & 0.39546 & 41.5 & 4.4 & sap\\
101616298 & 13.651 & 292.49025 & -0.03199 & 28.3 &  & man\\
101617139 & 14.731 & 292.49202 & -0.22664 & 35.7 & (4.1) & sap\\
101619414 & 13.126 & 292.49682 & -0.10495 & 23.8 & 3.2 & sap\\
101622423 & 14.531 & 292.50273 & -0.20858 & 31.2 & (4.0) & sap\\
101622447 & 13.513 & 292.50279 & 0.45704 & 50.4 & 5.4 & sap\\
101623741 & 14.873 & 292.50546 & 0.51612 & 27.5 & (3.5) & man\\
101624626 & 13.411 & 292.50721 & 0.00984 & 42.9 & (4.7) & sap\\
101626655 & 13.721 & 292.51126 & -0.16312 & 28.8 &  & man\\
101627819 & 14.621 & 292.51362 & 0.17405 & 27.1 &  & man\\
101628552 & 12.791 & 292.51520 & -0.27208 & 33.7 & (3.7) & sap\\
101629794 & 13.856 & 292.51781 & -0.10947 & 26.5 &  & man\\
101634748 & 13.561 & 292.52832 & 0.02923 & 31.1 & 3.7 & man\\
101634822 & 14.966 & 292.52848 & 0.14526 & 24.9 & (3.4) & man\\
101635594 & 13.493 & 292.53015 & 0.41928 & 26.7 & 3.2 & sap\\
101636040 & 15.211 & 292.53111 & 0.00377 & 35.7 &  & man\\
101638419 & 13.191 & 292.53611 & 0.02765 & 31.9 & 3.9 & sap\\
101641463 & 13.933 & 292.54243 & 0.45391 & 28.3 & 3.5 & man\\
101642089 & 13.681 & 292.54376 & 0.21842 & 29.4 & 3.9 & sap\\
101645783 & 13.056 & 292.55145 & -0.10843 & 22.7 &  & sap\\
101649216 & 12.461 & 292.55934 & -0.12285 & 19.4 & (2.5) & sap\\
101650702 & 13.931 & 292.56421 & -0.08882 & 12.6 & 1.9 & sap\\
101650995 & 15.061 & 292.56561 & -0.23266 & 25.0 &  & man\\
101654204 & 13.586 & 292.57313 & -0.24215 & 22.7 &  & sap\\
101661981 & 14.084 & 292.58634 & 0.09157 & 25.9 & 2.8 & man\\
101663605 & 13.471 & 292.58911 & 0.01487 & 26.1 & (3.5) & sap\\
101665008 & 13.556 & 292.59157 & 0.31293 & 62.2 & 6.2 & man\\
101681840 & 13.861 & 292.62024 & 0.08071 & 27.1 & (3.6) & man\\
101686314 & 13.926 & 292.62820 & 0.18007 & 26.0 &  & sap\\
101692807 & 13.611 & 292.63977 & -0.19429 & 12.4 & (1.8) & man\\
101701086 & 13.266 & 292.65412 & 0.12279 & 26.5 &  & sap\\
101708276 & 13.528 & 292.66683 & 0.05302 & 32.7 & (3.8) & sap\\
101737168 & 14.161 & 292.71810 & -0.01920 & 46.2 & 4.8 & man\\
101737628 & 14.271 & 292.71886 & 0.05476 & 27.1 & (3.6) & man\\
101741393 & 14.184 & 292.72543 & -0.12764 & 19.7 & (2.7) & man\\
101748322 & 12.624 & 292.73779 & -0.13966 & 52.6 & 5.3 & sap\\
101753179 & 13.976 & 292.74625 & -0.09557 & 24.7 & (3.3) & man\\
101758645 & 14.386 & 292.75604 & -0.04209 & 30.8 & 3.8 & man\\
110567305 & 14.681 & 290.98111 & 1.25287 & 52.3 & 5.3 & man\\
110569726 & 14.381 & 291.06331 & 1.01500 & 33.5 &  & man\\
110635144 & 14.788 & 290.94869 & 0.82256 & 35.0 &  & sap\\
110637723 & 12.714 & 291.03788 & 0.79087 & 21.4 & (2.8) & man\\
110639075 & 12.747 & 291.11563 & 0.77863 & 36.3 & 4.3 & sap\\
110649535 & 12.769 & 292.05737 & 1.49732 & 78.9 & 7.9 & sap\\
110651396 & 13.492 & 292.16917 & 0.51249 & 23.5 & 2.7 & man\\
\hline
\end{longtable}

\end{document}